%% file: paper-tycho.tex
\documentclass[iop, apj, numberedappendix]{emulateapj}

\usepackage{etoolbox}
\newtoggle{manuscript}
\togglefalse{manuscript}  % Set FALSE if using emulateapj or 2-col layout

\shorttitle{Tycho's Synchrotron Rims (\today)}  % <~ 44 char
\shortauthors{Tran et al. (\today)}  % Max three
\slugcomment{Accepted, ApJ, September 2, 2015}

\usepackage{amsmath}
\usepackage{CJK}  % aas.org/authors/author-names-non-roman-alphabets
\usepackage{booktabs}
\usepackage{hyperref}

\newcommand*{\mt}{\mathrm}
\newcommand*{\unit}[1]{\;\mt{#1}}  % vemod.net/typesetting-units-in-latex
\newcommand*{\abt}{\mathord{\sim}} % tex.stackexchange.com/q/55701
\newcommand*{\ptl}{\partial}

\newcommand*{\tsup}{\textsuperscript}

\usepackage{color}
\usepackage{soul}

\defcitealias{ressler2014}{R14}
\newcommand*{\Chandra}{\textit{Chandra}\ }
\newcommand*{\tsynch}{\tau_{\mt{synch}}}
\newcommand*{\mE}{m_\mt{E}}
\newcommand*{\Ecut}{E_{\mt{cut}}}
\newcommand*{\Bmin}{B_{\mt{min}}}
\newcommand*{\muG}{\unit{\mu G}}

\begin{document}
\begin{CJK}{UTF8}{bsmi}  % bsmi: traditional, gbsn: simplified

\title{Energy Dependence of Synchrotron X-Ray Rims in Tycho's Supernova
Remnant}

\author{
%Aaron Tran \altaffilmark{1,4},
Aaron Tran (陳宏裕)\altaffilmark{1,4},
Brian J. Williams\altaffilmark{1,5},
Robert Petre\altaffilmark{1},
Sean M. Ressler\altaffilmark{2},
Stephen P. Reynolds\altaffilmark{3}
}

\affil{
\tsup{1}X-ray Astrophysics Laboratory, NASA/GSFC, Code 662, Greenbelt, MD
20771, USA \\
\tsup{2}Dept. Physics, University of California, Berkeley, CA 94720, USA \\
\tsup{3}Dept. Physics, North Carolina State University, Raleigh, NC 27695, USA
}

\altaffiltext{4}{CRESST/University of Maryland, College Park, MD 20742}
\altaffiltext{5}{CRESST/Universities Space Research Association}

\begin{abstract}
Several young supernova remnants exhibit thin X-ray bright rims of synchrotron
radiation at their forward shocks.
Thin rims require strong magnetic field amplification beyond simple
shock compression if rim widths are only limited by
electron energy losses.
But, magnetic field damping behind the shock could produce similarly thin rims
with less extreme field amplification.
Variation of rim width with energy may thus discriminate between
competing influences on rim widths.
We measured rim widths around Tycho's supernova remnant in
5 energy bands using an archival $750 \unit{ks}$ \Chandra observation.
Rims narrow with increasing energy and are well described by either
loss-limited or damped scenarios,
so X-ray rim width-energy dependence does not uniquely specify a
model.
But, radio counterparts to thin rims are not loss-limited and better
reflect magnetic field structure.
Joint radio and X-ray modeling favors magnetic damping in Tycho's SNR with
damping lengths $\abt1$--$5$\% of remnant
radius and magnetic field strengths $\abt 50$--$400 \muG$ assuming Bohm diffusion.
X-ray rim widths are $\abt1$\% of remnant radius, somewhat smaller than
inferred damping lengths.
Electron energy losses are important in all models of X-ray rims,
suggesting that the distinction between loss-limited and damped models is
blurred in soft X-rays.
All loss-limited and damping models require magnetic fields $\gtrsim 20 \muG$,
affirming the necessity of magnetic field amplification beyond simple
compression.
\end{abstract}

% Quick ref: max rim width is ~9", min width is ~1.2" (of measurements)
% tycho radius ~ 4' = 240"
% Rims span 0.5-4% of remnant radius
% Typical widths are 2 to 3", around 0.8% to 1.3%.

% Six keywords, alphabetical order

% ============
% Introduction
% ============
\section{Introduction} \label{sec:intro}

% What are shock rims and how are they formed?  Why are they interesting?
Electrons accelerated in the forward shocks of young supernova remnants (SNRs)
emit synchrotron radiation strongly in the shock's immediate wake at radio
wavelengths and sometimes in X-rays.
In a few cases, they quickly turn off downstream, producing a shell-like
morphology of bright X-ray and radio rims/filaments due to line-of-sight
projection \citep{bamba2003,reynoso1997}.
Strong and time-variable synchrotron radiation \citep[e.g.,][]{uchiyama2007,
patnaude2007}, in conjunction with multiwavelength spectral modeling
\citep{aharonian2004, acero2010, ackermann2013}, suggests that electrons are
accelerated to TeV energies in young SNRs.
Although synchrotron emission due to accelerated electrons does not require
acceleration of an unseen hadronic component, the prevailing theory of
diffusive shock acceleration (DSA) should operate on both positive ions and
electrons.
Efficient hadron acceleration in supernova remnant shocks is a prime candidate
source for galactic cosmic rays up to the cosmic ray spectrum's ``knee'' at
around 3 PeV \citep{vink2012}.
But many fundamental questions about shock acceleration remain unanswered.
Under what conditions do shocks accelerate particles efficiently?
How are magnetic fields amplified in such shocks?
\citet{reynolds2008} reviews relevant observations and open questions to date.
These questions are relevant to many astrophysical settings, such as Earth's
bow shock \citep{ellison1990}, starburst galaxies \citep{heckman1990}, jets of
active galactic nuclei \citep{chen2014}, galaxy clusters
\citep{van-weeren2010}, and cosmological shocks \citep{ryu2008}.

% What have we learned so far, to 1st order?  Main inference is B ~100 \muG
% Introduce two key scenarios
Spectral and spatial measurements of synchrotron rims can constrain downstream
magnetic field strength and structure.
If rim widths are set by electron energy losses, post-shock magnetic fields
must be amplified to $\abt 10^2 \muG$ to account for the thinness of observed
rims \citep{vink2003, bamba2003, bamba2005-hist, volk2005, parizot2006}.
In these models, the magnetic field is assumed advected downstream and nearly
constant over rim widths.
Alternately, the magnetic field strength may be damped downstream of the shock
and prevent electrons from radiating efficiently, so that thin rims reflect
magnetic field variation rather than efficient particle acceleration and
synchrotron cooling \citep{pohl2005}.
Damping, in particular, may permit less extreme magnetic field amplification.
We refer to these as ``loss-limited'' and ``damped'' models for rim widths.
We shall see that models can range continuously between these two cases, and
that the distinction between the two can vary with observing frequency.

% Testing damping -- theory, spectra stuff
The possibility of damping in SNR shocks has not been fully tested.
\citet{marcowith2010} compared physically motivated magnetic damping models
to X-ray rim widths and synchrotron spectrum cut-offs and thus suggested that
only young SNRs (age $\lesssim 500 \unit{yr}$) can exhibit magnetic damping in
conjunction with efficient particle acceleration.
\citet{rettig2012} gave model predictions for several historical SNRs and
proposed discrimination based on filament spectra -- the expectation is that
damped spectra are softer, loss-limited harder.

% Testing damping -- direct modeling of shock rims
Hydrodynamic models can reproduce X-ray rim profiles reasonably well with both
loss-limited and damped magnetic field models
\citep{cassam-chenai2007, morlino2012, slane2014}.
However, loss-limited models generally cannot reproduce thin radio rims.
Radio-emitting GeV electrons do not lose substantial energy via radiation, so
modeled intensities rise to a broad maximum toward the remnant interior, then
gradually drop due to sphericity effects as the density drops in the interior.
\citet{reynolds1988} empirically modeled radio rims in the remnant of SN 1006
and concluded that such thin rims required sharp gradients in electron energy
density (from, e.g., time-variable particle acceleration) or magnetic field
strength.
\citet{cassam-chenai2007} used a 1-D hydrodynamic model with nonlinear DSA to
jointly model radio and X-ray rims in Tycho's supernova remnant; neither
loss-limited nor damped models could match radio profiles and radio/X-ray
intensities simultaneously.
Moreover, gradual X-ray spectral variation observed downstream of the shock front was
poorly reproduced with both models \citep{cassam-chenai2007}.
But damped models were able generate limb-brightened radio rims at the
forward shock, even if morphology was not entirely consistent with observation.
\citet{cassam-chenai2007} thus suggested that some combination of amplification
and magnetic field variation might explain radio morphology.

\begin{figure}
    \centering
    \plotone{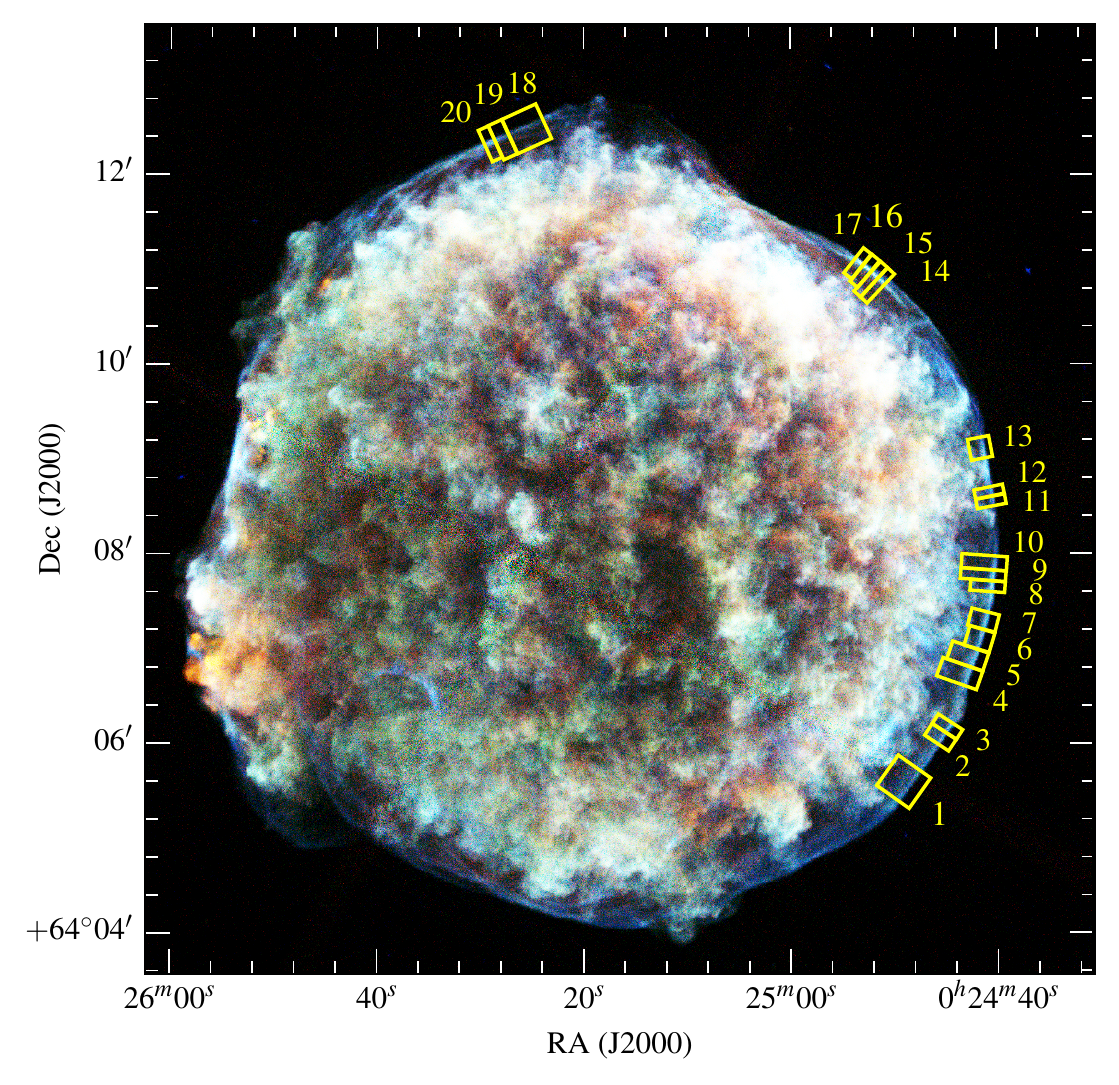}
    %\plotone{fig_snr-inv.png}  % Save ink when printing
    \caption{RGB image of Tycho with region selections overlaid.  Image bands
    are 0.7--1 keV (red), 1--2 keV (green) and 2--7 keV (blue).  Bold region
    labels (1, 16) indicate region selections shown in
    Figures~\ref{fig:spec},~\ref{fig:profiles}.  Filament 1: Regions 1--3,
    filament 2: regions 4--10, filament 3: regions 11-13, filament 4:
    regions 14--17, filament 5, regions 18--20.
    \label{fig:snr}}
\end{figure}

% Testing damping -- energy dependence.
Recently, \citet{ressler2014} (hereafter, \citetalias{ressler2014}) sought to
discriminate between damped and loss-limited rims by measuring rim width-energy
dependence in X-ray energies in SN 1006.
In the simplest models, rim widths are expected to be roughly energy-independent if rims are damped,
whereas widths should narrow with increasing energy if rims are energy
loss-limited. \citetalias{ressler2014} included a variety of effects which can blur this distinction.
To further test these models, we follow \citetalias{ressler2014} by measuring
X-ray rim widths at multiple energies in Tycho's supernova remnant (hereafter,
Tycho).
Tycho exhibits an extensive shell of synchrotron-dominated thin rims around its
periphery (Figure~\ref{fig:snr}); the rims show very little thermal emission,
consistent with expansion into a low density ISM \citep{williams2013}.
A deep $750 \unit{ks}$ exposure of the entire remnant from 2009 allows fine
sampling of the remnant rims.
We build upon previous estimates of magnetic field strength and particle
diffusion in Tycho that draw from multiwavelength observations and various
assumptions on CR acceleration \citep[e.g.,]{volk2002, volk2005, parizot2006,
morlino2012, rettig2012}.

% Paper roadmap
Our procedure closely follows that of \citetalias{ressler2014}.
We first review the model of \citetalias{ressler2014} used to model rim
profiles and widths, then describe the procedure for selecting, measuring, and
fitting rim widths to model width-energy dependence.
We explore degeneracies in model fitting and consider radio rim morphology as
an additional discriminant between models.

\section{Nonthermal rim modeling}\label{sec:models}

\subsection{Particle transport}\label{sec:transport}

% Prime reader, introduce injected spectrum
The energy and space distribution of electrons at a supernova remnant's forward
shock controls the synchrotron rims we see in X-ray and radio.
We assume that diffusive shock acceleration generates a power law distribution
of electrons with an exponential cut-off at the forward shock and model 1-D
steady-state plane advection and diffusion of the electron distribution
$f(E,x)$, where $E$ is electron energy and $x$ is distance downstream of the
forward shock:
\begin{equation} \label{eq:model}
    v_d \frac{\ptl f}{\ptl x}
    - \frac{\ptl}{\ptl x} \left(D\frac{\ptl f}{\ptl x}\right)
    - \frac{\ptl}{\ptl E} \left(bB^2E^2f\right)
    = K_0 E^{-s} e^{-E/\Ecut} \delta(x) ,
\end{equation}
following \citet{berezhko2004, cassam-chenai2007, morlino2010, rettig2012}.
The forward shock is located at $x=0$ with $x>0$ increasing downstream of the
shock; $D$ is the diffusion coefficient, $v_d$ is fluid velocity downstream of
the shock, and the constant $b \equiv 4e^4/9 m_e^4 c^7 = 1.57 \times 10^{-3}$
in appropriate CGS units arises from synchrotron power loss (i.e., $\ptl E/\ptl
t = -b B^2 E^2$, averaged over pitch angles).
The initial electron distribution is specified by an arbitrary normalization
$K_0$, DSA cut-off energy $\Ecut$ (given in
Section~\ref{sec:diffcoeff}), and spectral index $s = 2\alpha+1$.
% Condon and Ransom, ERA, (5C5)
\citet{zirakashvili2007} derive an electron energy spectrum with
super-exponential cut-off $e^{-(p/p_{\mt{cut}})^2}$, but we use a simple
exponential cut-off for simplicity and consistency with
\citetalias{ressler2014}.
We have not yet specified the cause of the injected spectrum cut-off
(see Section~\ref{sec:ecut}), but the functional form of a power law with
exponential cut-off is a good approximation to predictions for DSA spectra
limited by synchrotron losses, remnant age, or particle escape
\citep{webb1984, reynolds1998, reynolds2008}.
All constants and equations are given in CGS (Gaussian) units.  Our
presentation is somewhat abbreviated, but a fuller exposition and literature
review are given by \citetalias{ressler2014}.

% Tycho parameters
For Tycho, we adopted radio spectral index $\alpha = 0.58$ \citep{sun2011} and
hence electron spectral index $s = 2\alpha + 1 = 2.16$.  We assumed a remnant
distance $3 \unit{kpc}$ \citep[but cf.][]{hayato2010}, which gives a shock radius
of $1.08 \times 10^{19} \unit{cm}$ from the observed angular radius $240\arcsec$
\citep{green2014} and further sets shock velocities for each rim profile we
consider.  Tycho's forward shock velocity varies with azimuth by up to a factor
of 2 \citep{katsuda2008}; we linearly interpolated velocities reported by
\citet{williams2013} (rescaled to $3 \unit{kpc}$) to estimate individual shock
velocities for each region.
We assume a compression ratio of $4$ as for a strong shock (unmodified by
cosmic-ray pressure), and take downstream velocities $v_d$ to be one-fourth the
interpolated shock velocities.

% Assumptions, limitations of our model
We assume isotropic diffusion and only consider particle transport downstream
of the forward shock.
The velocity is assumed constant, as is the magnetic field for loss-limited
model rims, in contrast to the expected Sedov-Taylor similarity solution for
velocity and density in an adiabatic blastwave.
Our assumptions of constant velocity, magnetic field, and plane flow should be
reasonable as we generally consider synchrotron emission within 10\% of the
shock radius, $r_s$, from the forward shock, though a few models are followed
far enough inward that this approximation begins to break down.
We consider profile emission strictly upstream of both the contact
discontinuity and reverse shock \citep[cf.][]{warren2005}, and avoid regions
where the contact discontinuity overruns the forward shock.
Our modeling neglects flow and electron spectrum modification due to the nearby
contact discontinuity and particle acceleration at the reverse shock.
These effects could matter (and we do not quantify their importance), but we
expect that at Tycho's age, X-ray emission is dominated by forward shock
transport.
Given the uncertainty in shock morphology, our model seeks only to capture the
most relevant physics.
More sophisticated work may treat, e.g., sphericity, shock precursors,
anisotropic diffusion, and injection/acceleration efficiency \citep[e.g.,][and
references therein]{reville2013, bykov2014, ferrand2014}.

% Explain how we get model rims
To determine rim profiles and widths, we compute the electron distribution
using Green's function solutions by \citet{lerche1980} and \citet{rettig2012},
with the caveat that $D(x) B^2(x)$ is assumed constant; we discuss this
assumption further below.
The solutions are fully described in \citetalias{ressler2014} using notation
similar to ours.
The electron distribution may be integrated over the one-particle synchrotron
emissivity $G(y)$ to obtain the ``total'' emissivity:
\begin{equation} \label{eq:emissivity}
    j_{\nu}(x) \propto \int_0^\infty G(y) f(E,x) dE
\end{equation}
where $y \equiv \nu/(c_1 E^2 B)$ is a scaled synchrotron frequency and
$G(y) = y \int_y^\infty K_{5/3}(z) dz$ with $K_{5/3}(z)$ a modified Bessel
function of the second kind \citep{pacholczyk1970};
the constant $c_1 = 6.27 \times 10^{18}$ in CGS units.
Integrating emissivity over lines of sight for a spherical remnant yields
intensity as a function of radial coordinate $r$:
\begin{equation} \label{eq:intensity}
    I_{\nu}(r) = 2 \int_0^{\sqrt{r_s^2 - r^2}}
                   j_{\nu} \left( r_s - \sqrt{s^2 + r^2} \right) ds
\end{equation}
where $s$ is the line-of-sight coordinate and $r_s$ is shock radius.
We take the full width at half maximum (FWHM) of the resulting intensity
profile as our metric for modeled rim widths.  Using FWHM as opposed to, e.g.,
full width at three-quarters maximum, excludes measured rims and model
parameters where X-ray intensity does not drop to half maximum immediately
behind the rim.  Our results thus focus on the most well-defined rims in Tycho
rather than the global shock structure.

\subsection{Magnetic fields and damping}

% Introduce loss-limited vs. magnetic field
We consider two scenarios for post-shock magnetic field: (1) a constant field
$B(x) = B_0$ corresponding to loss-limited rims, and (2) an
exponentially damped field of form:
\begin{equation} \label{eq:bdamp}
    B(x) = \left(B_0 - \Bmin\right) \exp\left(-x / a_b\right)
           + \Bmin
\end{equation}
following \citep{pohl2005}.
Here $B_0$ is the magnetic field immediately downstream of the shock, i.e. $B_0
= B(x=0)$, and $a_b$ is an $e$-folding damping lengthscale; our use of $B_0$
for downstream magnetic field departs from typical notation.
A typical lengthscale for $a_b$ is $10^{16}$ to $10^{17} \unit{cm}$
\citep{pohl2005}, corresponding to $\abt 0.1$--$1\%$ of Tycho's radius.
Hereafter, we report $a_b$ in units of shock radius $r_s$ unless otherwise
stated.

% Explain the continuum of behavior, and how distinguishing damping really
% depends on both a_b and FWHMs (and hence l_ad, l_diff, B_0)
The distinction between damped and loss-limited rims is somewhat arbitrary; as
$a_b \to \infty$, model results converge to loss-limited rims.
Moreover, rims much thinner than the damping length are effectively
loss-limited as electrons radiate in a nearly constant magnetic field.
Furthermore, the dependence on observing frequency of electron losses and
diffusion means that a model may be damping limited at one frequency and
loss-limited at another.
In the following analysis, we deem all fits with finite $a_b$ to be damped, but
we compare rim widths and damping lengths from such fits further below to
better distinguish damped and loss-limited rim behavior, set by a combination
of $a_b$, $B_0$, and other model parameters.

% Diffusion
\subsection{Diffusion coefficient} \label{sec:diffcoeff}

% Bohm, Bohm-like, and MHD/turbulence-inspired diffusion
Most previous work has assumed Bohm-like diffusion in plasma downstream of SNR
shocks.  Bohm diffusion assumes that the particle mean free path $\lambda$ is
equal to the gyroradius $r_g = E/(eB)$, yielding diffusion coefficient
$D_{\mt{B}} = \lambda c / 3 = c E / (3 e B)$; here $c$ is the speed of light,
$E$ is particle energy, $e$ is the elementary charge, and $B$ is magnetic
field.  Bohm-like diffusion encapsulates diffusion scalings of $D \propto E$,
introducing a free prefactor $\eta$ such that $\lambda = \eta r_g$ allows for
varying diffusion strength.  However, Bohm diffusion at $\eta = 1$ is commonly
considered a lower limit on the diffusion coefficient at all energies.

% Define our diffusion coefficient, introduce and fix fiducial energy
We consider a generalized diffusion coefficient with arbitrary
power law dependence upon energy following, e.g., \citet{parizot2006}:
\begin{equation} \label{eq:diffcoeff}
    D(E) = \frac{\eta c E^\mu}{3 e B}
         = \eta_h D_{\mt{B}}\left(E_h\right) \left(\frac{E}{E_h}\right)^\mu
\end{equation}
where $\mu$ parameterizes diffusion-energy scaling and $\eta$ now has units
of $\mt{erg}^{1-\mu}$.
The right-hand side of equation~\eqref{eq:diffcoeff} introduces $\eta_h$, a
dimensionless diffusion coefficient scaled to the Bohm value at a fiducial
particle energy $E_h$.
Note that $\eta_h$ and $\eta$ are related as $\eta = \eta_h (E_h)^{1-\mu}$, and
$\eta = \eta_h$ for Bohm-like diffusion ($\mu = 1$).
For subsequent analysis, we take fiducial electron energy $E_2 = E_h$
corresponding to a $2 \unit{keV}$ synchrotron photon and report results in
terms of $\eta_2 = \eta_h$.
Although $E_2$ varies with magnetic field as $E_2 \propto B^{-1/2}$ and thus
$\eta_2$ may vary around Tycho's shock for $\mu \neq 1$, tying $\eta_h$ to a
fixed observation energy gives a convenient sense of diffusion strength
regardless of the underlying electron energies.

% Assumption needed for equation solutions
The solutions to equation~\eqref{eq:model} given by \citet{lerche1980} assume
$D(x) B^2(x)$ constant to render equation~\eqref{eq:model} semi-analytically
tractable.
Although this assumption has no obvious physical basis, it contains the
qualitatively correct behavior $D$ constant if $B$ is constant and $D$ smaller
for larger $B$.
We enforce it by modifying the diffusion coefficient in the damping model as,
following \citet{rettig2012}:
\begin{equation} \label{eq:ddamp}
    D(E,x) = \frac{\eta c E^\mu}{3 e B_0}
             \left[ \frac{\Bmin}{B_0} +
                    \frac{B_0 - \Bmin}{B_0} e^{-x/a_b} \right]^{-2} .
\end{equation}
This strengthens the spatial-dependence of the diffusion coefficient as
compared to the expected $D(x) \propto 1/B(x)$.

\subsection{Electron energy cut-off} \label{sec:ecut}

% e- cut-off energy result
We assume that the DSA process is limited by synchrotron losses at high
energies and hence determine the $\Ecut$ by equating synchrotron loss and
diffusive acceleration timescales.
Here we ignore the possibility of age-limited or escape-limited
acceleration \citep[e.g.,][]{reynolds1998}.  In the former case, low
magnetic-field strengths could mean that Tycho's age is less than a synchrotron
loss time; the maximum energy is obtained by equating the remnant age and
acceleration timescale.  In the latter case, the diffusion coefficient upstream
may increase substantially above some electron energy due to an absence of
appropriate MHD waves.  However, the magnetic field strengths we find below
justify the assumption of loss-limited acceleration.
For low energies and small synchrotron losses (cooling time longer than
acceleration time), electrons are efficiently accelerated; near or above the
cut-off energy, electrons will radiate or escape too rapidly to be accelerated
to higher energies and the energy spectrum drops off steeply.
The cut-off energy is given as:
\begin{align} \label{eq:ecut}
    \Ecut =
        &\left(8.3\unit{TeV}\right)^{2/(1+\mu)}
        \left(\frac{B_0}{100 \muG}\right)^{-1/(1+\mu)} \nonumber \\
        &\times \left(\frac{v_s}{10^8 \unit{cm\;s^{-1}}}\right)^{2/(1+\mu)}
        \eta^{-1 / (1+\mu)} .
\end{align}
This result is derived by \citet{parizot2006} for $\mu=1$ assuming a strong
shock with compression ratio $4$ and isotropic magnetic turbulence both
upstream and downstream of the shock.

% DSA (and our equations, Ecut, etc) depends on spatial variation of B / D.
The cut-off energy and accelerated electron spectrum depend on the
magnetic field, which varies in a damped model.
\citet{marcowith2010} show that various turbulent damping mechanisms can modify
the accelerated spectrum, as particles traveling downstream may not be
effectively reflected back across the shock and further accelerated.
Nevertheless, we make the simplifying assumption that particle acceleration is
controlled by diffusion at the shock and neglect spatially varying diffusion
and magnetic fields in the acceleration process; equation~\eqref{eq:ecut}
stands as evaluated with shock magnetic field strength $B_0$.
Cut-off energies of $\abt 1$--$10 \unit{TeV}$ can be plausibly achieved in the
presence of damping due to Alfv\'{e}n and magneto-sonic cascades
\citep{marcowith2010}.

% srcut constraint on diffusion
As the DSA imposed electron cut-off results in a cut-off of SNR synchrotron
flux, the synchrotron cut-off frequency $\nu_{\mt{cut}} = c_m E_{\mt{cut}}^2 B$
with $c_m = 1.82 \times 10^{18}$ in cgs units \citep[e.g.][]{pacholczyk1970},
which is the peak frequency emitted by electrons of energy $E_{\mt{cut}}$,
provides an independent observable to estimate shock diffusion and is given by:
\begin{align} \label{eq:cutoff}
    \nu_{\mt{cut}} =
        {}&c_m \left(13.3 \unit{erg}\right)^{\frac{4}{1+\mu}}
        \left(100 \muG\right)
        \left(2657 \unit{erg^2}\right)^{-\frac{1-\mu}{1+\mu}} \nonumber \\
        &\times \left( \frac{v_s}{10^8 \unit{cm/s}} \right)^{\frac{4}{1+\mu}}
        \left( \eta_2 \right)^{-\frac{2}{1+\mu}} .
    \end{align} % \nu_cut ={}&c_m ... enforces spacing
The cut-off frequency is independent of magnetic field $B$ for all values of
$\mu$ (but recall that the electron energies associated with $\eta_2$ will
depend on $B$ for $\mu \neq 1$).
\citet{parizot2006} previously used measurements of synchrotron cut-offs to
estimate diffusion coefficients in Tycho and other historical supernova
remnants.

% Discussion on synchrotron cooling
We point out that the electron spectrum softens downstream of the shock due to
synchrotron losses, but the local spectrum at any radial position will be a
steeply cut-off power law \citep{webb1984, reynolds1998}.
No steepening from $E^{-s}$ to $E^{-(s+1)}$ is observed because the cut-off
limits the electron spectrum at high energies.
A homogeneous source of age $t$ in which electrons are continuously accelerated
throughout, with an initial straight power law distribution to infinite energy,
will produce a steepened power law distribution above the energy at which the
synchrotron loss time equals the acceleration time \citep{kardashev1962}.
If we model emission without an initial exponential cut-off (i.e., inject a
straight power law) in a constant magnetic field, then integrated spectra in
our model would steepen by about one power.
But, those assumptions do not apply to the current situation of continuous
advection of electrons, with an energy distribution which is already a cut-off
power law, through a region of non-constant magnetic field.
See \citet{reynolds2009} for a fuller discussion of synchrotron losses in
non-homogeneous sources.

\subsection{Rim width-energy dependence} \label{sec:energydep}

\begin{figure*}

    \iftoggle{manuscript}{
        \epsscale{0.8}  % Artificially resized to fit caption+fig on ms page
        \plotone{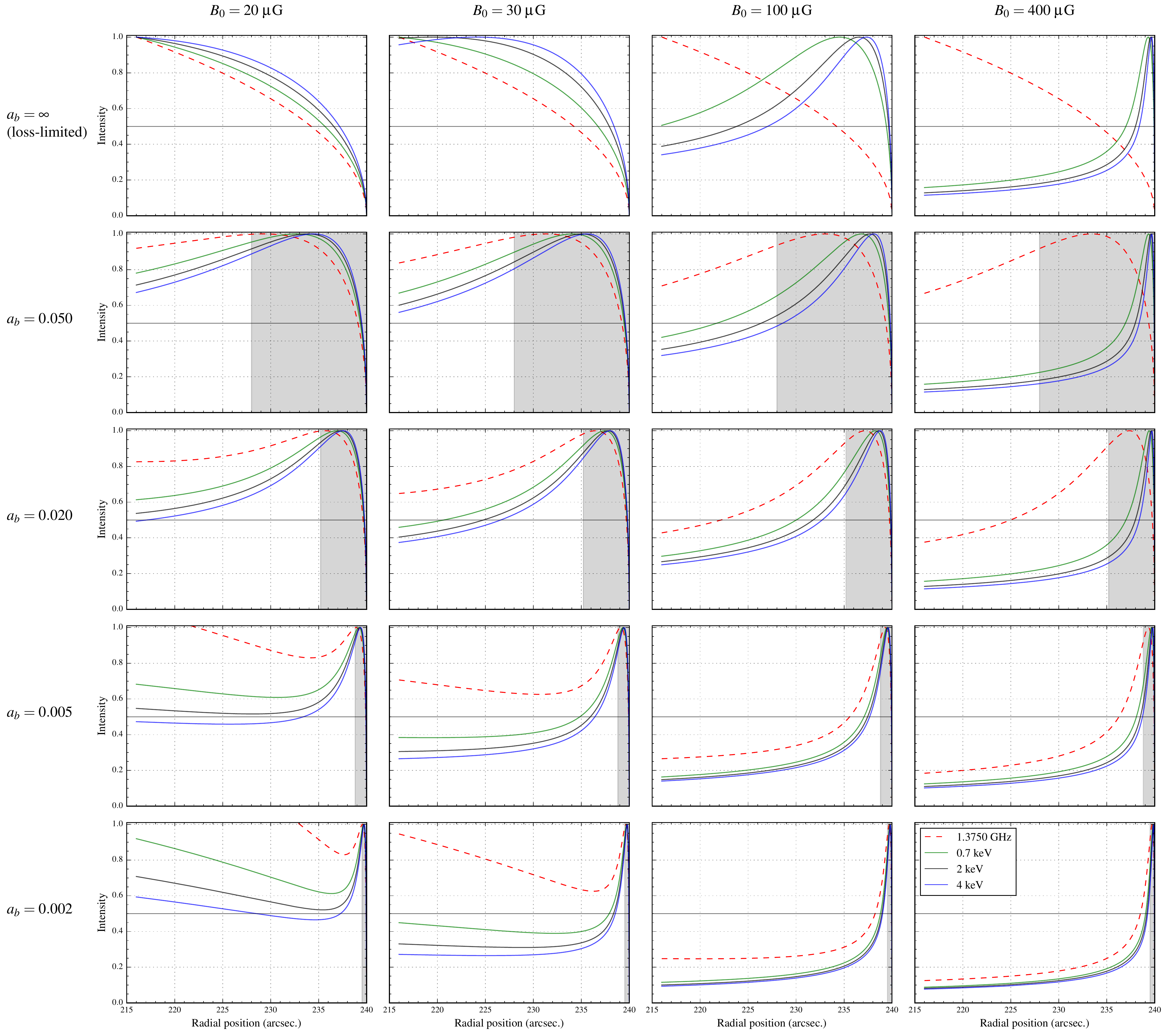}
        \epsscale{1}
    }{
        \plotone{fig_pargrid-eta2_1-model.pdf}
    }
    \caption{
        Radio and X-ray rim profiles for a range of magnetic fields
        ($B_0$) and damping lengths ($a_b$, shaded regions)
        with $\mu=1$, $\eta_2=1$, and $\Bmin = 5 \muG$ fixed.
        X-ray rim energies (0.7, 2, 4 keV) are representative of energy
        bands used in our rim width measurements.
        This plot summarizes several key features of our model:
        (1) radio profiles (dashed red) are strongly affected by damping for all
        values of $B_0$ because synchrotron losses are negligible at low
        electron energies.
        (2) X-ray profiles (solid green, black, blue) are influenced by
        synchrotron losses even in the presence of strong damping, which can be
        seen going from left to right (increasing $B_0$).
        (3) When X-ray rim widths are smaller than $a_b$, field damping only
        weakly affects rim widths and width-energy dependence (top right
        panels).
        (4) Strongly damped X-ray rims can show significant width-energy
        dependence in our model due to synchrotron losses beyond $a_b$. This
        occurs for smaller $B_0$ where the contrast between $B_0$ and $\Bmin$
        is less extreme (bottom left panels).
    }
    \label{fig:pargrid}
\end{figure*}

% Lay out three controlling mechanisms, prime reader w/ simple explanation
% Overlaps with more quantitative explanations below
X-ray rim widths are controlled by synchrotron losses, particle transport, and
magnetic fields immediately downstream of the shock, each of which influences
rim width-energy scaling differently.
If diffusion is negligible and the downstream magnetic field is constant,
loss-limited rims narrow with increasing energy as more energetic electrons
radiate and cool more quickly.  But diffusion will dilute this effect:
more energetic electrons may diffuse further upstream or downstream than would
be expected from pure advection, smearing out rims and weakening energy
dependence at higher energies.
Magnetic fields damped on a length scale comparable to filament widths
should also weaken rim width-energy dependence -- if the magnetic field turns
off, synchrotron radiation turns off regardless of electron energy.
Additionally, once $B_0$ varies, one observes electrons of different energy at
different distances behind the shock.

Figure~\ref{fig:pargrid} plots model X-ray and radio
profiles for a range of $B_0$ and $a_b$ values to illustrate how damping and
magnetic field strength impact width-energy dependence, which we will now
explore.  We discuss and incorporate model radio profiles into our analysis in
Section~\ref{sec:radio}.

\subsubsection{Undamped models}

% Define \mE
Following \citetalias{ressler2014}, we parameterize rim width-energy dependence
in terms of a scaling exponent $\mE$ defined as:
\begin{equation}
    w(\nu) \propto \nu^{\mE}
\end{equation}
where $w(\nu)$ is filament FWHM as a function of observed photon frequency
$\nu$, and the exponent $\mE = \mE(\nu)$ is energy dependent.
We may refer to observed photons interchangeably by energy or frequency $\nu$,
but $E$ is reserved for electron energy.

% lengthscale relations for loss-limited rims. R14 covers this in great depth.
To better intuit the effects of advection and diffusion on rim widths, we
introduce advective and diffusive lengthscales for bulk electron transport.
These depend on electron energy; we write them in terms of the peak
frequency radiated by electrons of energy $E$, $\nu = c_m E^2 B$.
\begin{equation} \label{eq:lad}
    l_{\mt{ad}} = v_d \tsynch
                \propto v_d B_0^{-3/2} \nu^{-1/2}
              % = \frac{v_d \sqrt{c_m}}{b} B_0^{-3/2} \nu^{-1/2}
\end{equation}
\begin{equation} \label{eq:ldiff}
    l_{\mt{diff}} = \sqrt{D \tsynch}
                  \propto \eta^{1/2} B_0^{-(\mu+5)/4} \nu^{(\mu-1)/4}
                % = (c/(3 e b))^{1/2} c_m^{-(\mu-1)/4} \eta^{1/2}
                  % B_0^{-(\mu+5)/4} \nu^{(\mu-1)/4}
\end{equation}
The characteristic time is the synchrotron cooling time
$\tsynch = 1 / (b B^2 E)$ with $b = 1.57 \times 10^{-3}$.
% b = 1.57e-3 given earlier, but redefined for convenience.
For $\mu = 1$, $l_{\mt{diff}}$ is independent of $\nu$ and both $l_{\mt{diff}}$
and $l_{\mt{ad}}$ scale as $B_0^{-3/2}$.
If both diffusion and magnetic field damping are negligible and electrons are
only loss-limited as they advect downstream, $\mE$ attains a minimum value
$\mE = -1/2$ as rim widths are set by $l_{\mt{ad}} > l_{\mt{diff}}$.
At higher energies where $l_{\mt{diff}} > l_{\mt{ad}}$, diffusion increases
$\mE$ from $-1/2$ to a value between $-1/4$ and $1/4$ for $\mu = 0$ and $2$
respectively \citepalias[Figure 3]{ressler2014}.
The presence of an electron energy cut-off decreases $\mE$ slightly in all
cases due to the decreased number of electrons and hence thinner rims at higher
energies \citepalias[Figure 5]{ressler2014}, but the qualitative behavior
is the same.

\subsubsection{Field damping effects}
\label{sec:energydep-damp}

% Intuition: how damping affects rim widths
% But, foreshadow and segue into explanation of energy-dep with damping
We expect magnetic damping to produce comparatively energy-independent rim
widths.
If synchrotron rim widths are set by magnetic damping at some observation
energy, then rims will be damped at all lower observation energies as well.
Then rim widths will be relatively constant (small $|\mE|$) below a threshold
energy and may decrease, or even increase once advection and/or diffusion
control rim widths at higher photon energies (advection: $l_\mt{ad} < a_b$;
diffusion: $l_{\mt{diff}} > l_{\mt{ad}}, a_b$).
Thus, we intuit that rim widths should roughly scale as
$w \sim \min\left( a_b, \max\left( l_{\mt{ad}}, l_{\mt{diff}} \right) \right)$.
This is correct except for one key region of parameter space: strong
damping with weak magnetic field amplification, where synchrotron losses
downstream of the FWHM create energy dependent widths even when $a_b \ll
l_{\mt{ad}} , l_{\mt{diff}}$.

% New text to explain losses in damped field (Sean)
This counter-intuitive energy dependence for strongly damped models
occurs when
rim brightness remains above half-maximum within $\abt a_b$ of the
shock. Farther downstream, synchrotron losses in the reduced magnetic field
$\abt \Bmin$ drive intensity down to half-maximum; losses in a nearly
constant field cause energy dependence despite magnetic damping at the shock.
Synchrotron losses far from the shock will only affect the spectrum at
higher frequencies; at lower frequencies, the spectrum will be effectively
constant after the magnetic field's initial decay. Thus, there will be a
characteristic frequency below which the FWHM ceases to be defined. The energy
dependence of rim FWHMs becomes large and diverges near this frequency.
The characteristic frequency and value of $\mE$ are sensitive to our definition
of rim width (FWHM), but the physical behavior we describe (divergence of $\mE$
at characteristic frequency, for appropriate damping parameters) will occur
regardless of our definition of rim width.

% ============
% Observations
% ============
\section{Observations}
\label{sec:observations}

% Spectrum figure pushed way up to help 2-column layout...
\begin{figure*}[]
    \plotone{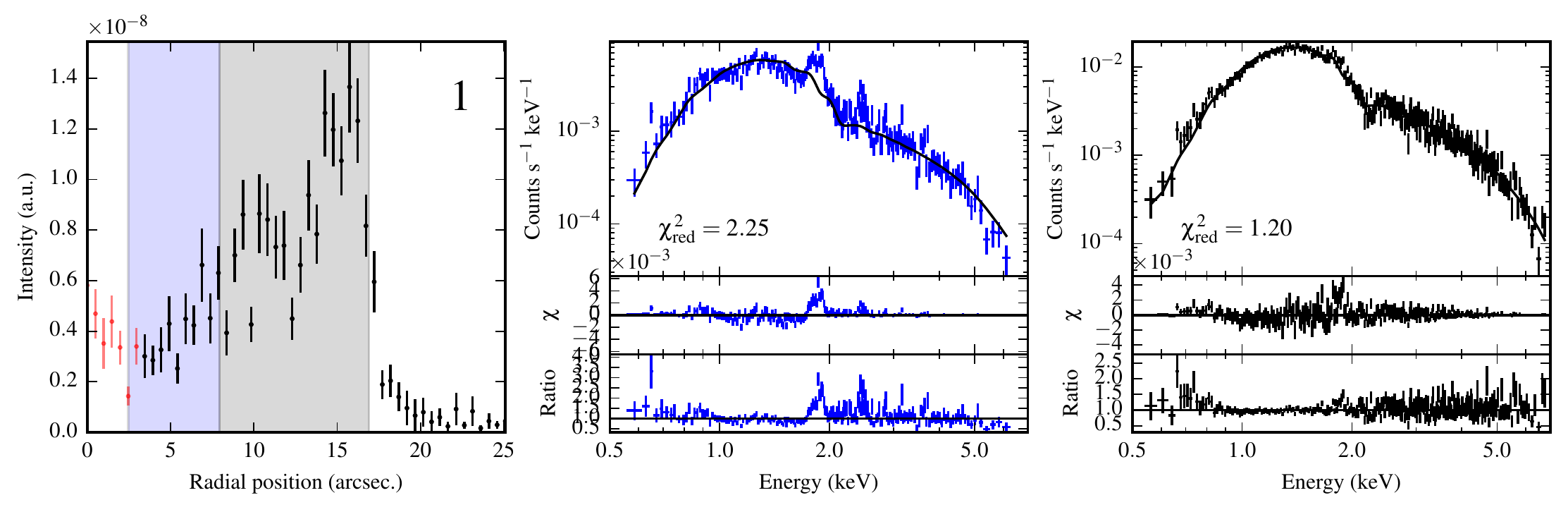} \\
    \plotone{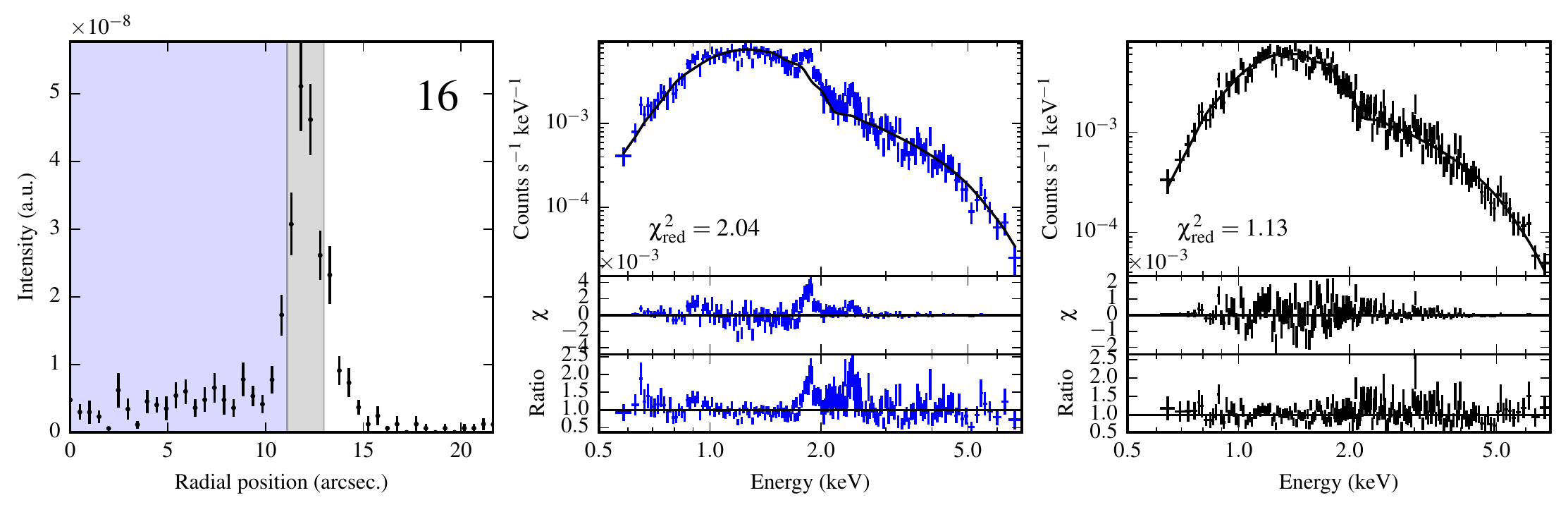}
    \caption{Spectra and fits from Regions 1 (top) and 16 (bottom) show varying
        rim morphology; Region 1 shows a rim where the $0.7$--$1 \unit{keV}$
        peak could not be fit.  Left: $4.5$--$7 \unit{keV}$ profiles with
        downstream (blue) and rim (grey) sections highlighted.  Intensity is in
        arbitrary units (a.u.).  Middle: downstream spectra with absorbed
        power law fit; Si and S lines at $1.85$, $2.45 \unit{keV}$ are clearly
        visible.  Right: rim spectra with absorbed power law fit show that
        rims in each region are likely free of thermal line emission.}
    \label{fig:spec}
\end{figure*}

\subsection{Data and region selections}
\label{sec:regions}

We measured synchrotron rim full widths at half maximum (FWHMs) from an
archival \Chandra ACIS-I observation of Tycho
(RA: 00\tsup{h}25\tsup{m}19\fs0, dec: +64\arcdeg08\arcmin10\farcs0; J2000)
between 2009 Apr 11 and 2009 May 5 (PI: J. Hughes;
\dataset[ADS/Sa.CXO\#obs/10093--10097]{ObsIDs: 10093--10097},
\dataset[ADS/Sa.CXO\#obs/10902--10906]{10902--10906}); \citet{eriksen2011}
present additional observation information.
The total exposure time was $734 \unit{ks}$.
Level 1 \Chandra data were reprocessed with CIAO 4.6 and CALDB 4.6.1.1 and kept
unbinned with ACIS spatial resolution $0.492\arcsec$.
Merged and corrected events were divided into five energy bands:
0.7--1 keV, 1--1.7 keV, 2--3 keV, 3--4.5 keV, and 4.5--7 keV.
We excluded the 1.7--2 keV energy range to avoid \ion{Si}{13} (He$\alpha$)
emission prevalent in the remnant's thermal ejecta which might contaminate our
nonthermal profile measurements.

We selected 20 regions for profile extraction around Tycho's shock
(Figure~\ref{fig:snr}) based on the following criteria: (1) filaments should be
clear of spatial plumes of thermal ejecta in \Chandra images, which rules out,
e.g., areas of strong thermal emission on Tycho's eastern limb; (2) filaments
should be singular and localized, so multiple filaments should either not
overlap or completely overlap; (3) filament peaks should be evident above the
background signal or downstream thermal emission (rules out faint southern
filaments).
We accepted several regions with poor quality peaks in the lowest energy band
($0.7$--$1 \unit{keV}$) so long as peaks in all higher energy bands were clear
and well-fit.
We grouped regions into 5 filaments by visual inspection of the remnant.
Within each filament, we chose region widths to obtain comparable counts at the
thin rim peak.
All measured rim widths are at least $1\arcsec$.
The narrowest rim widths may be slightly over-estimated due to the \Chandra
point-spread function (PSF) at $\abt4\arcmin$ off the optical axis, which has
FWHM $\leq 1.4\arcsec$ at $6.4 \unit{keV}$ and $\leq 1\arcsec$ at $1.5
\unit{keV}$.
For simplicity, we neglect PSF effects in our analysis.
Our observations of rim width-energy dependence are thus somewhat conservative
at the highest energy

% ===== Supplemental information =====

% Chandra Cycle 17 POG, Figure 4.13 gives 50% energy enclosed radii 4' off-axis
% for each ACIS-I chip, blurring + \Chandra aspect error included.
% I estimate values from figure w/ ~0.1 arcsec accuracy
% 50\% encircled energy radius is <= 1" at 1.5 keV, <= 1.4" at 6.4 keV.
%
% HRMA/ACIS-I3: y2 = 2.674, y1 = 2.863    HRMA/ACIS-I2: y2 = 2.675, y1 = 2.867
% 2.845 @ 6.4 keV, 50\%   = 1.1"          2.831 @ 6.4 keV, 50\%   = 1.2"
% 2.894 @ 1.49 keV, 50\%  = 0.8"          2.889 @ 1.49 keV, 50\%  = 0.9"
%
% HRMA/ACIS-I1: y2 = 6.179, y1 = 6.373    HRMA/ACIS-I0: y2 = 6.179, y1 = 6.369
% 6.333 @ 6.4 keV, 50\%   = 1.2"          6.300 @ 6.4 keV, 50\%   = 1.4"
% 6.378 @ 1.49 keV, 50\%  = 1"            6.369 @ 1.49 keV, 50\% = 1"
%
% HRMA/HRC-I: y2 = 6.039, y1 = 6.244
% 6.11 @ 6.4 keV, 50\%    = 1.65"
% 6.18 @ 1.49 keV, 50\%   = 1.3"

% ----------------
% Filament spectra
% ----------------

\subsection{Filament spectra}
\label{sec:spec}

% Filament spectra figure pushed way up to help 2-column layout

We extracted spectra at and immediately behind thin rims in each region
(``rim'', ``downstream'' spectra respectively) to confirm that rim width
measurements are not contaminated by thermal line emission.
The two extraction regions are determined by our empirical fits of rim profile
shape (Section~\ref{sec:fwhms}).  The rim section is the smallest sub-region
containing the measured FWHM bounds from all energy bands.
The downstream section extends from the interior thin rim FWHM to the intensity
minimum behind the rim (specifically, the downstream profile fit domain bound
described in Section~\ref{sec:fwhms}).
To illustrate our selections, Figure~\ref{fig:spec} plots example rim profiles
($4.5$--$7 \unit{keV}$) with the downstream and rim sections highlighted.

% Spectrum data, binning, background
Spectra were binned to a minimum of 15 cts/bin.  We extracted
background spectra from circular regions (radius $\abt 30\arcsec$) around the
remnant's exterior; for each region's rim and downstream spectra, we subtracted
the closest background region's spectrum.
% Background region size is valid for data-tycho/bkg-2/ selections

% Upstream spectra and numbers
We fit each region's rim and downstream spectra to an absorbed power law model
(XSPEC 12.8.1, \texttt{phabs*powerlaw}) between $0.5$--$7 \unit{keV}$ with
photon index $\Gamma$, hydrogen column density $N_{\mt{H}}$, and a
normalization as free parameters.  Table~\ref{tab:spec} lists best fit
parameters and reduced $\chi^2$ values for all regions.  Rim spectra are
well-fit by the power law model alone; the best fit photon indices ($2.4$--$3$)
and column densities ($0.6$--$0.8 \times 10^{22} \unit{cm^{-2}}$) are
consistent with previous spectral fits to Tycho's nonthermal rims
\citep{hwang2002, cassam-chenai2007}.

\begin{table*}
    \scriptsize
    \centering
    \caption{Absorbed power law spectrum fit parameters
        \label{tab:spec}}
    \input{tab-spec.tex}
    \tablecomments{Absorbed power law fit parameters are photon index $\Gamma$
        and hydrogen column density $N_{\mt{H}}$.
        \texttt{srcut} fits performed in log-frequency space; $h$ is Planck's
        constant and $\nu_{\mt{cut}}$ a cut-off frequency.
        Horizontal rules group individual regions into filaments.}
\end{table*}

% This table is still useful to look at the fit parameters (e.g., to make
% quantitative statements about excised/Gaussian line fit quality)
%\begin{table*}
%    \scriptsize
%    \centering
%    \caption{Region spectra fit parameters\label{tab:spec-pt2}}
%    \input{suppl-spec-pt2.tex}
%    \tablecomments{As described in text,
%    we consider absorbed power law fits with lines fitted to two Gaussians
%    (here characterized by equivalent width and line energy), or with lines
%    manually excised.  See comments on Table~\ref{tab:spec}.}
%\end{table*}

% Downstream spectra and numbers
Downstream spectra are poorly fit by the absorbed power law model due to
thermal contamination from \ion{Si}{13} and \ion{S}{15} He$\alpha$ line
emission at $1.85$ and $2.45$ keV.  To confirm that thermal emission is
dominated by these two lines near the shock, we also performed fits with (1)
both lines excised ($1.7$--$2.0 \unit{keV}$, $2.3$--$2.6 \unit{keV}$ counts
removed) and (2) with both lines fitted to Gaussian profiles.  Fits with lines
excised yield $\chi^2_{\mt{red}}$ values between $1$--$5$.  Fits with lines
fitted to Gaussian profiles yield $\chi^2_{\mt{red}}$ values $0.83$--$1.6$.  In
both fits (lines excised or modeled), we find somewhat smaller best fit column
densities ($0.3$--$0.8 \times 10^{22} \unit{cm^{-1}}$) but similar best fit
photon indices ($2.6$--$3.1$), compared to those of the rim spectra.  The
consistent photon indices indicate that the same synchrotron continuum is
present beneath thermal line emission.

% Addition: srcut fit description
We also fitted ``rim'' spectra (Section~\ref{sec:spec}) to the absorbed XSPEC
model \texttt{srcut}, modified to fit in log-frequency space.
\texttt{srcut} models a power law X-ray synchrotron spectrum set by a radio
spectral index $\alpha$ with an exponential cut-off parameterized by cut-off
frequency $\nu_{\mt{cut}}$ \citep{reynolds1998, reynolds1999}.
The fit values of $\nu_{\mt{cut}}$, in particular, permit an independent
estimate of $\eta_2$ from equation~\eqref{eq:cutoff}.
The radio spectral index is fixed to $\alpha = 0.58$ \citep{sun2011} as done in
our transport modeling, and we fit for absorption column density $N_{\mt{H}}$
and cut-off frequency $\nu_{\mt{cut}}$ in each region.
Table~\ref{tab:spec} lists best spectrum fit parameters for each
region.  The fitted cut-off frequency is typically $0.3 \unit{keV/h}$,
consistent with fits by \citet{hwang2002}, but Regions 11, 12 have unusually
high cut-off frequencies $0.55$, $0.88 \unit{keV/h}$ consistent with harder rim
spectra.

% Takeaway -- we are good to go
Our spectral fitting confirms that all selected region are practically free of
thermal line emission, as already suggested by visual inspection
(Figure~\ref{fig:snr}).  Excluding $1.7$--$2 \unit{keV}$ photons in rim
width measurements further limits thermal contamination as $1.85 \unit{keV}$ Si
line emission is over a third of Tycho's thermal flux as detected by \Chandra
\citep{hwang2002}.

% --------------------------
% FWHM measurement procedure
% --------------------------
\subsection{Filament width measurements}
\label{sec:fwhms}

\begin{figure*}
    \plotone{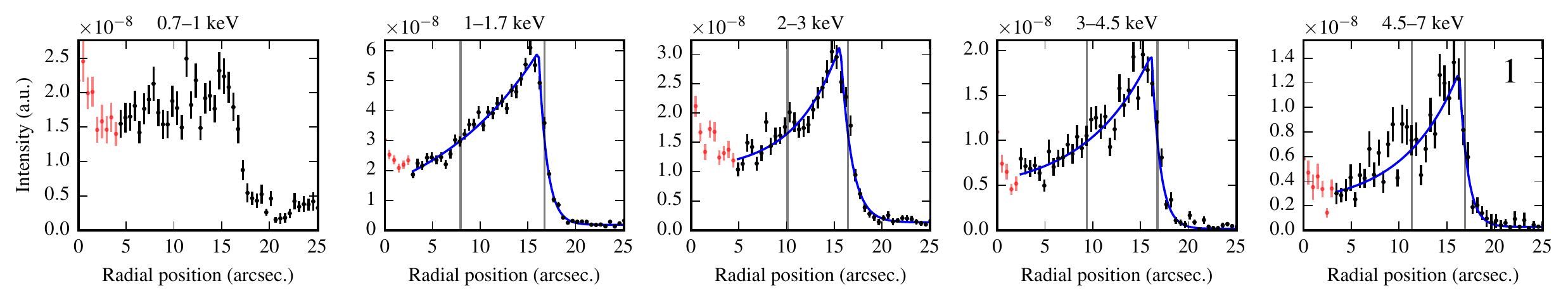} \\
    \plotone{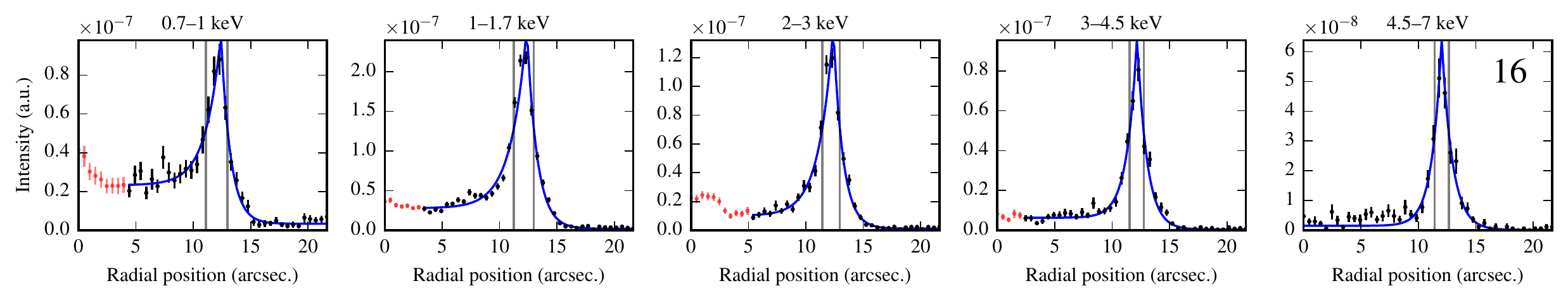}
    \caption{Best fit profiles with measured FWHMs demarcated for each energy
        band in Region 1 (top) and Region 16 (bottom).  Hereafter, we use
        Regions 1 and 16 to illustrate results for profiles of differing
        quality and absolute width.  We could not measure a $0.7$--$1
        \unit{keV}$ FWHM in Region 1, reflected in Table~\ref{tab:fwhms}.  Data
        in red were excluded from profile fit domains as described in text.}
    \label{fig:profiles}
\end{figure*}

We obtained radial intensity profiles in five energy bands from $\abt
10$--$20\arcsec$ behind the shock to $\abt 5$--$10\arcsec$ in front for each
region.  To increase signal-to-noise, we integrate along the shock
($5$--$23\arcsec$) in each region.  Plotted and fitted profiles are reported in
vignetting and exposure-corrected intensity units; error bars were computed from
raw counts assuming Poisson statistics.  Intensity profiles peak sharply within
$\abt 2$--$3\arcsec$ behind the shock, demarcating the thin rims, then fall off
gradually until thermal emission picks up further behind the shock.

We fitted rim profiles to a piecewise two-exponential model:
\begin{equation} \label{eq:prof}
    h(r) =
    \begin{cases}
        A_u \exp \left(\frac{r_0 - r}{w_u}\right) + C_u, &r \geq r_0 \\
        A_d \exp \left(\frac{r - r_0}{w_d}\right) + C_d, &r < r_0
    \end{cases}
\end{equation}
where $h(r)$ is profile height and $r$ is radial distance from remnant center.
The rim model has 6 free parameters: $A_u, r_0, w_u, w_d, C_u$, and $C_d$;
$A_d = A_u + (C_u - C_d)$ enforces continuity at $r=r_0$.
Our model is similar to that of \citet{bamba2003, bamba2005-hist} and differs
slightly from that of \citetalias{ressler2014}.
To fit only the nonthermal rim in each intensity profile, we selected the fit
domain for each profile as follows.  The downstream bound was set at the first
local data minimum downstream of the rim peak, identified by smoothing the
profiles with a 21-point ($\abt 10\arcsec$) Hanning window.  The upstream bound
was set at the profile's outer edge.  Figure~\ref{fig:profiles} illustrates the
fit domain selections for two example regions.

From the fitted profiles we extracted a full width at half maximum (FWHM) for
each region and each energy band after subtracting a constant background term
$\min(C_u, C_d)$.  We could not resolve a FWHM in 8 of 20 regions at 0.7--1 keV
(Table~\ref{tab:fwhms}); in these regions, either the downstream FWHM bound
would extend outside the fit domain or we could not find an acceptable fit to
equation~\eqref{eq:prof}.  We were able to resolve FWHMs for all regions at
higher energy bands (1--7 keV).

To estimate FWHM uncertainties, we horizontally stretched each best fit
profile by mapping radial coordinate $r$ to
$r'(r) = r (1 + \xi (r-r_0)/(50\arcsec-r_0))$ with $\xi$ an arbitrary stretch
parameter and $r_0$ the best fit rim center from equation~\eqref{eq:prof},
yielding a new profile $h'(r) = h(r'(r))$.
We varied $\xi$ (and hence rim FWHM) to vary each profile fit $\chi^2$ by 2.7
and took stretched FWHMs as upper/lower bounds on reported FWHMs.
This procedure again follows \citetalias{ressler2014}.

% -------------
% Model fitting
% -------------
\subsection{Filament model fitting}
\label{sec:fits}

% How we ran model fits (knobs, energy band mapping, errors, fit procedure)
We fit model FWHM predictions given by equation~\eqref{eq:model} to
measured rim widths as a function of energy by varying several physical
parameters: magnetic field strength $B_0$, normalized diffusion coefficient
$\eta_2$, diffusion-energy scaling exponent $\mu$, and minimum field strength
$\Bmin$ and lengthscale $a_b$ for a damped magnetic field.
We mapped each width measurement to the lower energy limit of its energy band;
e.g., $0.7$--$1 \unit{keV}$ is assigned to $0.7 \unit{keV}$ and fitted to model
profile widths at $0.7 \unit{keV}$.
Width errors in our least squares fits average the positive and negative errors
on each FWHM measurement.
For a given set of model parameters, we numerically computed intensity profiles
and hence model FWHMs as detailed in Section~\ref{sec:transport}; we then used
a Levenberg-Marquardt fitter to seek model parameters yielding best fit FWHMs.
To assist the nonlinear fitting, we tabulated model FWHM values on a large grid
of model parameters and used best fit grid parameters as initial guesses for
fitting.
We required $\eta_2$ to be positive and deemed best fit values with $\eta_2
\geq 10^5$ and $B_0 \geq 10 \unit{mG}$ to be effectively unconstrained.
In subsequent analysis we focus on fits with $\mu = 1$ and $\eta_2 = 1$ fixed,
though we discuss the effects of varying both $\mu$ and $\eta_2$ (and fitting
for $\eta_2$) as well (recall that for $\mu = 1$, $\eta$ is
energy-independent).

% How we twiddled model knobs
The purely loss-limited model (constant downstream magnetic field $B_0$) has
three parameters $\mu$, $\eta_2$, and $B_0$.
To make nonlinear fitting tractable, we fixed $\mu$ in all fits and considered
$\mu = 0$, $1/3$, $1/2$, $1$, $1.5$, and $2$.
In particular, nonlinear diffusion-energy scalings with $\mu = 1/3$ and $1/2$
may arise from Kolmogorov and Kraichnan turbulent energy spectra respectively
\citep{reynolds2004}.

% Damped case
For damped magnetic field fits, we held the remnant interior field strength
$\Bmin$ constant at $5 \muG$, slightly higher than typical intergalactic values
of $\abt 2$--$3 \muG$ \citep{lyne1989, han2006}.
We stepped $a_b$ through $14$ different values between $0.5$ and $0.002$
(sampling most finely between $0.01$ and $0.002$).
To ensure that damped fits generate rims influenced by magnetic damping, we
arbitrarily require that the rim FWHM at $2 \unit{keV}$ be strictly greater
than the fit value of $a_b$; we revisit this requirement below.
For best fits, we report the value of $a_b$ yielding the smallest $\chi^2$
value, with the caveat that that our $a_b$ sampling is relatively coarse.

% Model resolution error
Predicted rim widths are subject to resolution error in the numerical integrals
(discretization over radial coordinate, line-of-sight coordinate, electron
distribution, Green's function integrals).  We chose integration resolutions
such that the fractional error in model FWHMs associated with halving or
doubling each integration resolution is less than $1\%$ for the parameter space
relevant to our filaments.
The maximum resolution errors in a sample of parameter space are typically
$0.1$--$1\%$, but mean and median errors are typically an order of magnitude
smaller than maximum errors.

% =======================
% Results, FWHMs and fits
% =======================
\section{Results}

% -----------------
% FWHM measurements
% -----------------
\subsection{Rim widths}
\label{sec:fwhm-results}

% Present table of FWHM results
Measured rim widths decrease with energy in most regions and energy bands.
Table~\ref{tab:fwhms} reports FWHM measurements for all of our regions.
We also report $\mE$ values for all but the lowest energy band, computed
point-to-point between discrete energy bands as:
\begin{equation}
    \mE(E_2) = \frac{\ln(w_2/w_1)}{\ln(E_2/E_1)}
\end{equation}
where $w_1, w_2$ and $E_1, E_2$ are FWHMs and lower energy values for each
energy band -- e.g., $\mE$ at $1 \unit{keV}$ is computed using FWHMs from
$0.7$--$1 \unit{keV}$ and $1$--$1.7 \unit{keV}$, with $E_2 = 1 \unit{keV}$ and
$E_1 = 0.7 \unit{keV}$.  Errors on $\mE$ are propagated in quadrature from
adjacent FWHM measurements.

\begin{table*}
    \iftoggle{manuscript}{
        \tiny
    }{
        \scriptsize
    }
    \centering
    \caption{Measured full widths at half max (FWHMs) for all regions.
             \label{tab:fwhms}}
    \input{tab-fwhms.tex}
\end{table*}

% Observations of mE and rim widths, errors
Although the measurement scatter is quite large, shown dramatically in the
point-wise computed $\mE$ values, the mean rim width decreases consistently
with increasing energy.
Furthermore, mean $\mE$ values are consistently negative and tend smoothly
towards $0$ (weaker energy-dependence) with increasing energy.
Errors on FWHM measurements are typically $\lesssim 10\%$, reflecting the high
quality of the underlying \Chandra data.
Scatter in FWHM measurements may be attributed in part to (1) our measurement
procedure, which depends on an empirical choice of profile fit function, and
(2) variation in Tycho's rim morphology (e.g., Figure~\ref{fig:profiles}).

% -------------------------
% Model fit results, tables
% -------------------------
\subsection{Model fit results}
\label{sec:fit-results}

\begin{figure*}
    \centering
    \iftoggle{manuscript}{
        \epsscale{0.8}
        \plotone{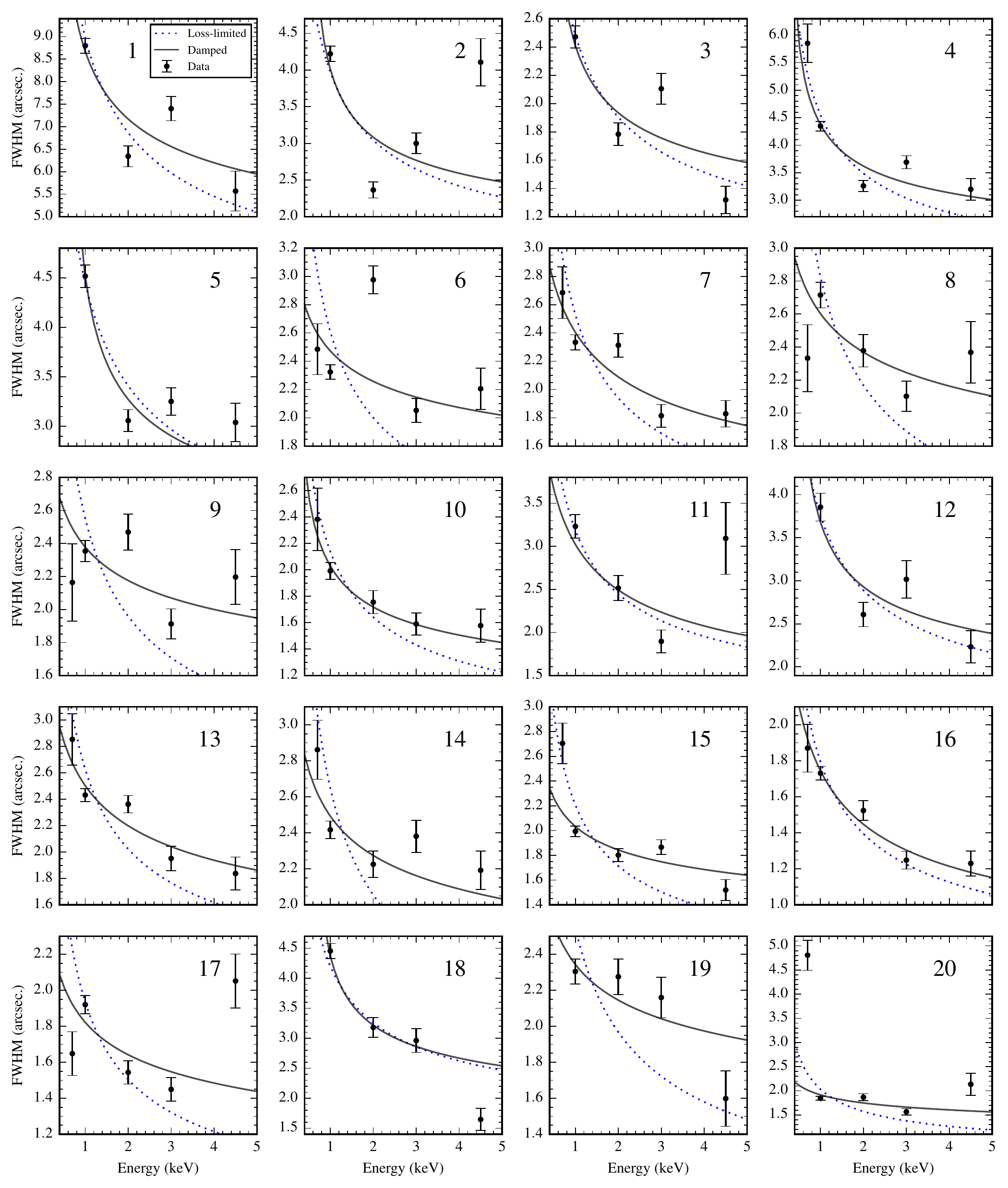}
        \epsscale{1}
    }{
        \plotone{fig_energywidth-subplot.pdf}
    }
    \caption{Rim width fits as a function of energy for loss-limited and
    damped fits with $\mu = 1$ and $\eta_2=1$ fixed for all regions, from
    best fit parameters in Table~\ref{tab:fits-all-eta2one}.
    Damped model predictions (e.g., Region 5) are not given at low energies if
    FWHMs cannot be calculated for model profiles (i.e., modeled intensity
    behind thin rim exceeds half-maximum of rim peak within model domain of
    $\abt 20\arcsec$).
    Ordinate ($y$) axis limits vary between subplots and are offset from the
    origin to better show model predictions and data variation.
    \label{fig:fits-all}}
\end{figure*}

% Introduce results (mu=1, eta2=1)
Best fit results for loss-limited and damped cases are summarized in
Figure~\ref{fig:fits-all} and Table~\ref{tab:fits-all-eta2one} with $\mu = 1$
and $\eta_2 = 1$ both fixed.
In all tables, we report values of $\chi^2$ rather than $\chi^2_\mt{red}$ to
ease comparison of different fits, as we often manually stepped parameters that
were held constant in fitting.

\begin{table}
    \scriptsize
    \centering
    \caption{Best width-energy fit parameters, $\mu = \eta_2 = 1$
        \label{tab:fits-all-eta2one}}
    \input{tab-fits-all-eta2one.tex}
    \tablecomments{Fits for Regions 1--3, 5, 11, 12, 18, and 19 have 3 degrees
    of freedom; all others have 4.
    The choice of a best $a_b$ value may be construed as removing one
    additional dof.
    Damped fits require $a_b$ to be smaller than the FWHM at
    $2\unit{keV}$ in order to rule out effectively loss-limited fits with
    large $a_b$.}
\end{table}

\begin{table}
    \scriptsize
    \centering
    \caption{Best model fits for all regions, $\eta_2$ derived from
        \texttt{srcut} fits, $\mu=1$
        \label{tab:fits-all-srcutlog}}
    \input{tab-fits-all-srcutlog.tex}
    \tablecomments{$\eta_2$ values are computed from equation~\eqref{eq:cutoff}
    and held fixed in model fits.
    All comments for Table~\ref{tab:fits-all-eta2one} apply to \texttt{srcut} fits
    as well.\newline\newline}
    % Avoid awkward widow/orphan in 2-col layout, on page w/ 3 tables
\end{table}

% srcutlog fits.  Effect of mu/eta_2 in loss-limited fits;
% mu variation tempered by varying eta2
Table~\ref{tab:fits-all-srcutlog} reports model fits with $\eta_2$ fixed to
values derived from \texttt{srcut} spectrum fits for $\nu_{\mt{cut}}$ and hence
$\eta_2$ from equation~\eqref{eq:cutoff}, with $B_0$ varying freely;
cutoff-derived $\eta_2$ values of $\abt 10$ require increased magnetic fields
compared to the $\eta_2 = 1$ case if rims are purely loss-limited.
We derive different $\eta_2$ values for varying $\mu$ and find that fitted
$B_0$ varies only weakly with $\mu$; for example, Regions 1 and 16 have best
fit values of $B_0$ increasing over $252$--$266 \muG$ and $660$--$700 \muG$
respectively as $\mu$ ranges between $0$--$2$, for loss-limited fits.

% Damping and loss-limited fits both plausible
Magnetically damped rims are able to fit width-energy dependence in our
measurements at least as well as purely loss-limited rims for an acceptable
range of $a_b$ values.
As loss-limited models are a subset of damping models, this is expected.
But fits with damping lengths small enough to exert influence on rim widths
are also permissible.
In both loss-limited and damped fits, we do not observe systematic
variation of fit parameters with azimuth around the remnant.

\subsection{Fitting parameter degeneracy}

We briefly explore the effects of varying $\eta_2$ and other fit
parameters, before focusing on results with $\mu = \eta_2 = 1$ for simplicity.
Fixing $\eta_2$ from measured $\nu_{\mt{cut}}$ values does not yield any
obvious insight.
But, diffusion coefficients computed in this manner may give more credible
derived estimates of $B_0$ if the DSA assumptions invoked are accurate.

% Degeneracy in loss-limited model
Loss-limited model values of $B_0$ and $\eta_2$ may covary without strongly
altering fit quality in many regions.
If diffusion is the primary control on rim width -- i.e.,
$l_{\mt{diff}} > l_{\mt{ad}}$ -- then $\eta_2$ and $B_0$ become degenerate as
the product $\eta^{1/2} B_0^{-3/2}$ exerts most control on rim width $w \sim
l_{\mt{diff}}$ (equation~\eqref{eq:ldiff}).
If advection is the primary control on rim width (widths narrow rapidly with
energy; i.e., $\mE \sim -0.5$), then $\eta_2 \ll 1$ becomes unimportant and
fits are well-behaved with effectively one free parameter.

% Effect of varying mu upon chi^2 and eta_2
We also performed loss-limited model fits with $\mu$ fixed between $0$
and $2$ and both $\eta_2$ and $B_0$ free.
Fits with $\mu \lesssim 1$ generally yield larger parameter values and errors
for both $\eta_2$ and $B_0$, and they are more likely to be ill-constrained
(e.g., $\eta_2 > 10^3$ or $B_0 > 10^3 \muG$).
Fits to the same data with varying $\mu$ can yield $\eta_2$ varying by $1$--$2$
orders of magnitude.
The $\chi^2$ values for individual fits are variable and large ($\gg 1$), so we
cannot favor or disfavor particular values of $\mu$ and $\eta_2$.
But, neglecting the magnitude of our $\chi^2$ values, values of $\mu \geq 1$
are qualitatively favored by $\chi^2$ in most regions.
This trend may be partially an artifact of the correlation between $B_0$ and
$\eta_2$ as they are not entirely independent parameters, but they are less
correlated for larger $\mu$; fits at smaller $\mu$ may have fewer (non-integer)
degrees of freedom, partially offsetting our observations.

% Then, degeneracy modification in the damping model.  eta_2 behavior...
Damping modifies the degeneracy in $B_0$ and $\eta_2$.
For small values of $a_b$ (strong damping), increased diffusion ($\eta_2$) can
cause rim widths at all energies to narrow counterintuitively.
Speculatively, spatial variation of the diffusion coefficient may oppose
downstream advection through a decreased effective velocity
$v_d - (\ptl D/\ptl x)$ in our transport equation:
\begin{align}
    \left[ v_d - \frac{\ptl D(x)}{\ptl x} \right] \frac{\ptl f}{\ptl x}
    - D(x) \frac{\ptl^2 f}{\ptl x^2}
    - \frac{\ptl}{\ptl E} \left(bB^2E^2f\right) \nonumber \\
    = K_0 E^{-s} e^{-E/\Ecut} \delta(x) ,
\end{align}
requiring thinner rims as $\eta_2$ and hence $\ptl D(x) / \ptl x$ increase.
Moreover, the effective velocity may impact rim widths even if advection
dominated.
In practice, if $\eta_2$ varies freely in damped fits, we observe that smaller
values of $a_b$ permit and favor smaller best fit values of both $B_0$ and
$\eta_2$.

% B0 estimates in loss-limited and damped models
Best fit $B_0$ values are smallest for $\eta_2$ approaching $0$ and $\mu = 2$;
intuitively, $\mu > 1$ strengthens diffusion at energies $>2\unit{keV}$ for
fixed $\eta_2$, permitting a smaller best fit $\eta_2$ and hence smaller $B_0$.
Fixing $\eta_2 = 1$ (Table~\ref{tab:fits-all-eta2one}), however, ties the range
of $B_0$ values to the range of observed rim widths.
Our minimum loss-limited values of $B_0$ are consistent with prior estimates of
$\abt 200$--$300 \muG$ for advection-dominated transport \citep{volk2005,
parizot2006, morlino2012}.
If loss-limited, Tycho's rims require strong magnetic field amplification to
$\abt 100\times$ typical galactic field values of $\abt 2$--$3 \muG$, versus
the expected $4\times$ amplification from a strong shock.

Magnetic damping fits permit much smaller values of $B_0$, as expected.
The minimum value of $B_0$ is around $20 \muG$, which would require no field
amplification beyond that from a strong shock.
Fixing $\Bmin = 2 \muG$ instead of $5 \muG$ permits smaller fit $B_0$ values,
though fit values for both $\eta_2$ and $B_0$ still display considerable
scatter.

% ==========
% Discussion
% ==========
\section{Discussion}

% ----------------
% Main null result
% ----------------
\subsection{Damping is poorly-constrained by X-ray width-energy dependence}
\label{sec:damp-fit-disc}

% Width-energy dependence is not so helpful
Width-energy dependence is not as sensitive a discriminant between damping and
loss-limited models as may be intuitively expected.
Both loss-limited and damped fits do not perfectly capture sharp drop-offs
(especially between $0.7$--$1 \unit{keV}$ and $1$--$2 \unit{keV}$), and the
data show large scatter (Figure~\ref{fig:fits-all}).
In most regions, best fit width-energy curves have $\mE \sim -0.2$.
Only sharp decreases in rim width ($\mE \sim -0.5$) can disfavor damping as a
control on rim widths (e.g., Region 18).

\begin{figure}
    \centering
    \iftoggle{manuscript}{
        \epsscale{0.5}
        \plotone{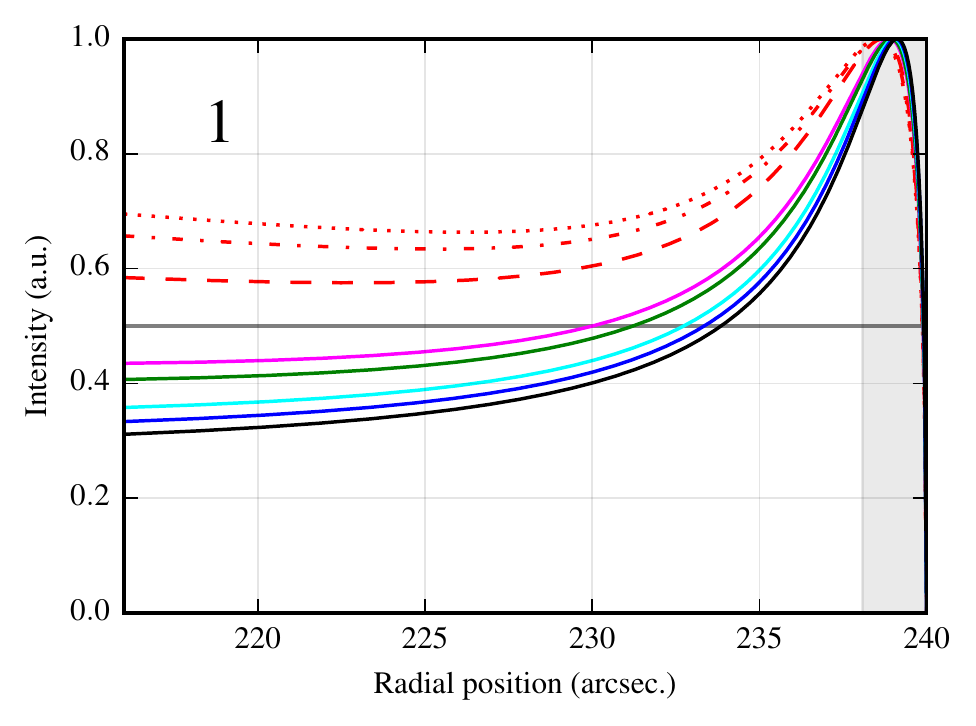} \\
        \plotone{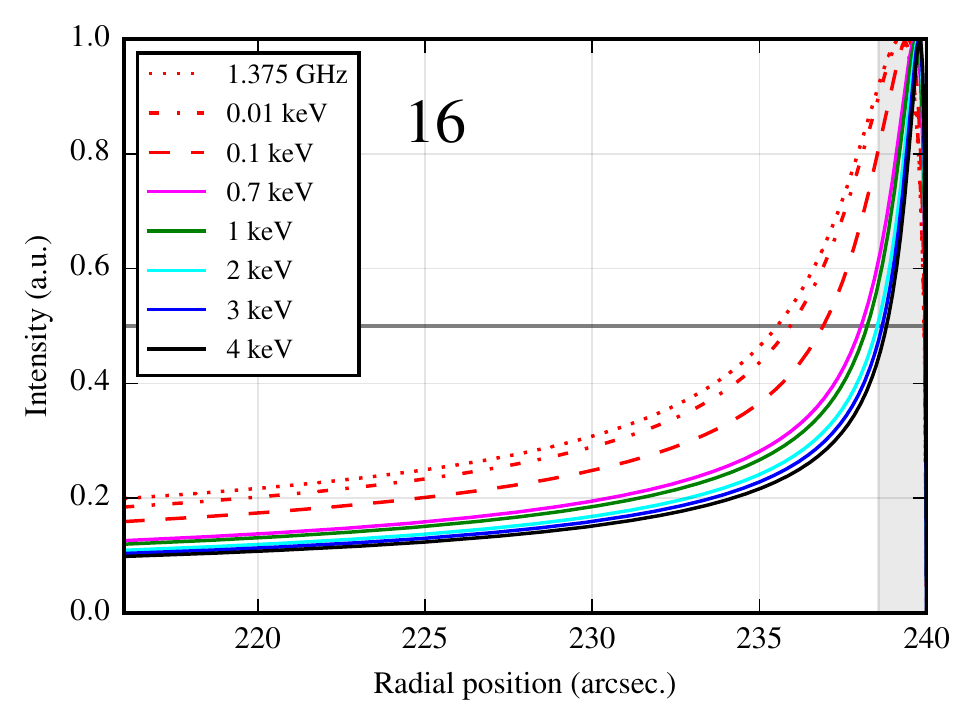}
        \epsscale{1}
    }{
        \plotone{fig_prfs-fit-damp-01.pdf} \\
        \plotone{fig_prfs-fit-damp-16.pdf}
    }
    \caption{Model predictions illustrate weak-field (Region 1) and
        strong-field damping (Region 16) for damped best fit parameters with
        $\mu=1$ and $\eta_2=1$ fixed.
        In X-ray energies ($0.7$--$4.5 \unit{keV}$) model profiles are not
        strongly energy dependent, but weak-field damping profiles evolve and
        show no measurable FWHM at sufficiently low energies.
        Profiles are normalized to peak thin rim intensity, and shaded regions
        indicate damping lengthscale $a_b$.
        Model parameters are given in Table~\ref{tab:fits-all-eta2one}.
        \label{fig:rims}}
\end{figure}

% Explain the dichotomy in energy dependence
When width-energy dependence is weak, both loss-limited and damped models give
best fits with similar profiles -- amplified magnetic fields in both models
cause rim intensities to drop sharply.
When width-energy dependence is stronger, damping profiles yield
energy dependent rims if magnetic fields are small (say,
$\lesssim 50 \muG$) and the damping lengthscale is much smaller than rim FWHMs.
Figure~\ref{fig:rims} shows this contrasting behavior
using best fit parameters for Regions 1 and 16.

% Weak-field damping
The best damped fit parameters for Regions 1--5, 10, 12, and 18 predict a
$1.375 \unit{GHz}$ radio profile that does not drop below 50\% of peak
intensity within $\abt 20\arcsec$ downstream of the shock, which we treat as
having no measurable FWHM.
%All other regions predict a measurable rim FWHM at $1.375 \unit{GHz}$.
These regions are all best fit with field $B_0 < 40 \muG$.
We refer to this model behavior as ``weak-field'' damping, associated with weak
magnetic fields and stronger emission intensity immediately downstream of the
thin rim.
As observation frequency decreases, rim and trough
contrast decreases, causing rim
FWHMs to increase and eventually become unmeasurable.
Rim width-energy dependence strengthens dramatically
($\mE \to -\infty$), permitting model fits to replicate strong width-energy
dependence in observed rim widths.
Weak-field fit parameters require energy losses downstream of the
shock to produce energy dependence in a damped field, as discussed
in Section~\ref{sec:energydep-damp}.

% Strong-field damping
Damping model fits to all other regions predict consistently thin rims with
measurable FWHMs at decreasing energy.
The width-energy dependence parameterized by $\mE$ trends
towards zero at low energy, indicating that rim widths are comparatively energy
independent.
At higher energy, such rims narrow slightly with increasing energy ($\mE \sim
-0.2$) to match the observed width-energy dependence in X-rays.
The gradual increase in $|\mE|$ with energy is expected in the damping model as
advection and/or diffusion take control of rim widths at increasing energy, as
discussed in Section~\ref{sec:energydep}.
The best damped fits for these regions all have larger $B_0$ values than in the
``weak-field'' damping case.
We refer to these fits as giving rise to ``strong-field'' damping, associated
with stronger magnetic fields, weaker emission intensity behind the thin rim,
and clear rims with measurable FWHMs at low photon energies (at and below soft
X-rays).

% Explanation and caveats
We are unable to determine whether weak- or strong-field damping is descriptive
of Tycho's shock magnetic field, given the large $\chi^2$ values on our model
fits.
Several regions may be well-fit by ``weak-field'' and ``strong-field'' damping
alike.
But, the qualitative behavior of $B_0$ is suggestive; very strong fields at the
shock ($\gtrsim 100 \muG$) with magnetic field damping on a scale
comparable to rim FWHMs appear incompatible with significant energy
dependence.
The distinction between weak- and strong-field damping may be equally well
described as strong versus weak damping.  The dichotomy reflects additional
degeneracy between $a_b$ and $B_0$ -- for a given rim width, increased damping
length $a_b$ requires increased $B_0$ to reproduce the same width, and vice
versa.

\begin{table*}
    \scriptsize
    \centering
    \caption{Lengthscale analysis
        \label{tab:length}}
    \input{tab-lengthscale.tex}
    \tablecomments{All lengthscales computed at fiducial energy $2 \unit{keV}$
    from best fits with $\eta_2 = 1$ and $\mu = 1$.
    Ratio of $l_{\mt{ad}}/l_{\mt{diff}}$ is same for loss-limited and damped
    fits and depends only on plasma velocity $v_d$ and observation energy, as
    $l_{\mt{ad}}/l_{\mt{diff}}$ is independent of $B_0$ for $\mu = 1$.}
\end{table*}

% Further discussion - lengthscale analysis, hybrid behaviour
Although model fits are poorly-constrained, we attempt to further explore how
damping and synchrotron losses set rim widths for damped fits with finite
$a_b$.
Table~\ref{tab:length} compares rim widths at $2 \unit{keV}$ to advection
lengthscale $l_{\mt{ad}}$ and damping lengthscale $a_b$ from the loss-limited
and damped fits of Table~\ref{tab:fits-all-eta2one}.
As discussed in Section~\ref{sec:energydep}, loss-limited rim widths are
qualitatively set by either damping or the dominant transport process as
$w \sim \min\left( a_b, \max\left( l_{\mt{ad}}, l_{\mt{diff}} \right) \right)$.
At $2 \unit{keV}$ with $\eta_2 = 1$ and $\mu = 1$ fixed, the ratio
$l_{\mt{ad}}/l_{\mt{diff}}$ is nearly constant; variation ($1.2$--$1.4$) arises
solely from azimuthal variation in shock velocity.
If rims are loss-limited and diffusion is negligible, we anticipate
rim widths $w = 4.6 l_{\mt{ad}}$, where the factor $4.6$ may be derived
assuming spherical symmetry and an exponential synchrotron emissivity
\citep{ballet2006}.  At $2 \unit{keV}$, we find that loss-limited rim widths
$w \sim 5 l_{\mt{ad}}$ due to diffusion.
Enforcing $w(2 \unit{keV}) / a_b > 1$ for damped fits still permits
$l_{\mt{ad}} / a_b < 1$ as $w(2\unit{keV}) \lesssim 5 l_{\mt{ad}}$, with damping
decreasing $w(2 \unit{keV})$ below the expected loss-limited width at a given
$B_0$.  We may impose a tighter or looser bound, but our results
should not be greatly affected given large uncertainty in our fits and
associated degeneracy between $B_0$ and $a_b$.

% SN 1006
Our modeling can also reassess the significance
of rim width-energy dependence in the remnant of SN 1006 presented by
\citetalias{ressler2014}.
We fit averaged filament widths in SN 1006 measured by \citetalias{ressler2014}
to our damping model using the same procedure as for Tycho.
We fix $\Bmin = 5 \muG$, although best fit magnetic fields for SN 1006 are
smaller than those of Tycho due to the much wider filaments of SN 1006.
A more extensive search of parameter space produces acceptable damping (hybrid)
models with more rapid shrinkage than found in \citetalias{ressler2014}.
Damped fits are comparable to or better than loss-limited fits in 3 of 5
filaments in SN 1006.
Fits to two filaments with strong energy dependence ($\mE \sim -0.5$) favor a
loss-limited model with sub-Bohm diffusion ($\eta_2 \ll 1$), though
sub-Bohm diffusion may be an unphysical result pointing to
oversimplifications in the model.
The width-energy dependence in SN 1006 ($\mE \sim -0.3$ to $-0.5$) is overall
slightly stronger than in Tycho, and the best damped fits for SN 1006 all fall
into the ``weak-field'' case.
The best fit $B_0$ values are less than $40 \muG$ in the damped model, compared
to $\abt 100$--$200 \muG$ in the loss-limited model.
If thin radio rims in SN 1006 \citep{reynolds1986}
indicate magnetic field damping on lengthscales comparable to X-ray
filament widths, some magnetic
damping is not incompatible with rim width narrowing.
A future study of SN 1006 with recent multi-frequency
Karl G. Jansky Very Large Array data (PI: D. Green) may
verify whether radio and X-ray rims can be jointly described by an amplified
and subsequently damped magnetic field.

% ------------------------------------------
% Model correctness - are our results valid?
% ------------------------------------------
%\subsection{Tycho distance estimate weakly affects model fits}
\subsection{Model assumptions and correctness}

% Distance estimates don't affect much; but we can give handy scalings
Our models adopted a distance $d$ to Tycho of $3 \unit{kpc}$, but estimates for
Tycho's distance range between $2.3$--$4 \unit{kpc}$ \citep{hayato2010}.  As a
larger remnant distance would increase both physical filament widths and shock
velocity estimates from proper motion, we may derive expected scalings for
modeled $\eta_2$ and $B_0$ as functions of assumed distance $d$.  The advective
lengthscale, equation~\eqref{eq:lad}, may be rearranged to obtain:
\begin{align}
    B_0 =\; &(3.17 \muG) \left(\frac{v_d}{10^8 \unit{cm/s}}\right)^{2/3}
                \nonumber \\
            &\times \left(\frac{l_{\mt{ad}}}{0.01 \unit{kpc}}\right)^{-2/3}
                \left(\frac{h\nu}{1 \unit{keV}}\right)^{-1/3}
\end{align}
or, more simply,
\begin{equation}
    B_0 \propto \left(v_d\right)^{2/3}
                \left(l_{\mt{ad}}\right)^{-2/3} \nu^{-1/3} .
\end{equation}
Both $l_{\mt{ad}}$ and $v_d$ scale linearly with remnant distance $d$ and thus
their effects cancel in determining the magnetic field.  If diffusion is the
primary control on filament lengthscales, equation~\eqref{eq:ldiff} yields:
\begin{equation}
    \eta_2 \propto \left(l_{\mt{diff}}\right)^2 B_0^{(\mu+5)/2} \nu^{-(\mu - 1)/2}
\end{equation}
Model fits with varying distance obey both scalings, $\eta_2
\propto d^2$ and $B_0$ constant.  When comparing model fits with remnant
distances of $3 \unit{kpc}$ and $4 \unit{kpc}$, the deviation from the
idealized scaling is $\lesssim 1 \%$ for $B_0$ and $\abt 1$--$5\%$ for
$\eta_2$.  As varying remnant distance $d$ leaves width-energy scaling $\mE$
invariant, the relative contributions of $l_{\mt{ad}}$ and $l_{\mt{diff}}$
should also be invariant.  Then both lengthscales should scale simultaneously
with $d$, yielding the observed behavior.  In the damped model, a larger
distance $d$ will require larger physical damping lengths from magnetic
turbulence.  But, fitted values of $a_b$ should remain unchanged as we report
$a_b$ in units of shock radius $r_s$.

% D(x) B^2(x) constant assumption doesn't affect core physics
The exponential variation in $D(x)$ (equation~\eqref{eq:ddamp}) may not be
physically reasonable; as noted above, the assumption $D(x)
\propto 1 / B^2(x)$ gives rise to sharp gradients in diffusion coefficient and
even inverts the effect of $\eta_2$ on our model profiles (where larger
$\eta_2$ can cause rim widths to narrow).  Nevertheless, the modeled behavior
appears physically reasonable.
Only X-ray profiles are impacted by the assumption on $D(x)$ as radio profiles
assume no diffusion.
And, model behavior driven by advection and magnetic damping --
namely, rim width-energy dependence in a damped magnetic field -- occurs for
$D=0$, and cannot be an artifact of our assumption that
$D(x) B^2(x)$ is constant.

% Data quality -- don't need more counts
We emphasize that additional counts from averaging measurements or selecting
larger regions will likely not improve our ability to constrain $B_0$ and
$\eta_2$ from width-energy modeling.
This is easily seen from Figure~\ref{fig:fits-all} as well as large
$\chi^2$ values in Tables~\ref{tab:fits-all-eta2one} that reflect relatively
tight errors in our FWHM profile measurements (Table~\ref{tab:fwhms}).

\subsection{Other potential constraints}

% Discussion on spectral indices + softening downstream of shock
% Emphasize that our approach is different from that of Cassam-Chenai+ 2007
Rim width-energy dependence is a morphological manifestation of spectral
softening downstream of the forward shock;
previously, \citet{cassam-chenai2007} also sought to distinguish loss-limited
and damped rims with a careful spectral study and 1-D hydrodynamical model.
Although knowledge of radial spectral variation, in principle, fully determines
rim profiles, there is not a clear relationship between observed spectral
variation and rim widths alone.
Our model, for example, replicates observed rim width variation but
underpredicts the observed spectral variation.
The more sophisticated model of \citet{cassam-chenai2007} similarly has trouble
reproducing the observed radial gradient in spectral photon index.
Photon indices predicted in our model, for integrated model rim spectra
similar to those of Figure~\ref{fig:spec} and Table~\ref{tab:spec}, are also
somewhat ill-constrained. Enforcing limits on acceptable model photon
indices \citep[e.g., Fig. 16][]{cassam-chenai2007}
could help constrain shock parameters or identify discrepancies
between model assumptions and measurements.

% New paragraph on relative radio/X-ray fluxes to address reviewer comment
Estimates of magnetic field strength that depend upon global models for Tycho's
evolution \citep[e.g.,][]{morlino2012} may not help constrain our rim model
results.
In such models, assumptions on particle acceleration and magnetic field
evolution downstream of the shock (e.g., further damping or amplification at
the reverse shock and contact discontinuity) would affect the spatially
integrated spectrum of Tycho.
We have focused on effects immediately behind the forward shock, making as few
assumptions as possible about particle acceleration and magnetic fields
throughout the remnant.
It may not make sense to extrapolate our results to estimate magnetic fields
throughout the remnant or, similarly, to use global models to constrain our
results given all assumptions involved.

% ----------------------------------------
% Constraining rims via radio measurements
% ----------------------------------------
\section{Joint radio and X-ray modeling}
\label{sec:radio}

% Introduction
Thin radio rims spatially coincident with X-ray synchrotron rims may help
constrain magnetic field damping.
Radio synchrotron emission in the remnant interior may arise from not only
recently shock-accelerated electrons but also long-lived electrons interacting
with a turbulent field inside the remnant, as $\abt$GeV electrons have cooling
times of order $10^5$--$10^7 \unit{yr}$ in $10$--$100 \muG$ magnetic fields.
But simple models for synchrotron emissivity as a function of density in Sedov
and pre-Sedov dynamical stages predict radio synchrotron profiles rising
gradually to broad maxima well inside the shock radius
\citep[e.g.][]{reynolds1981, reynolds1988}.
Our steady-state planar transport model predicts monotonically increasing
emission downstream of the forward shock in a loss-limited model, until
sphericity becomes important and our model is inapplicable.
On scales of a few percent of the remnant radius, only magnetic field damping
can cause emission to decrease.

% Previous work, and how we differ
The idea of modeling X-ray and radio profiles jointly was pioneered by
\citet{cassam-chenai2007}, who found that sharp radio rims in Tycho were not
reproduced by a loss-limited model.
\citet{cassam-chenai2007} considered loss-limited and damping profiles
consistent with physical constraints applied in a hydrodynamical model.
Here, we neglect physical constraints and estimate a damping length necessary
to generate radio rims with shape similar to observations.

\begin{figure}
    \centering
    \plotone{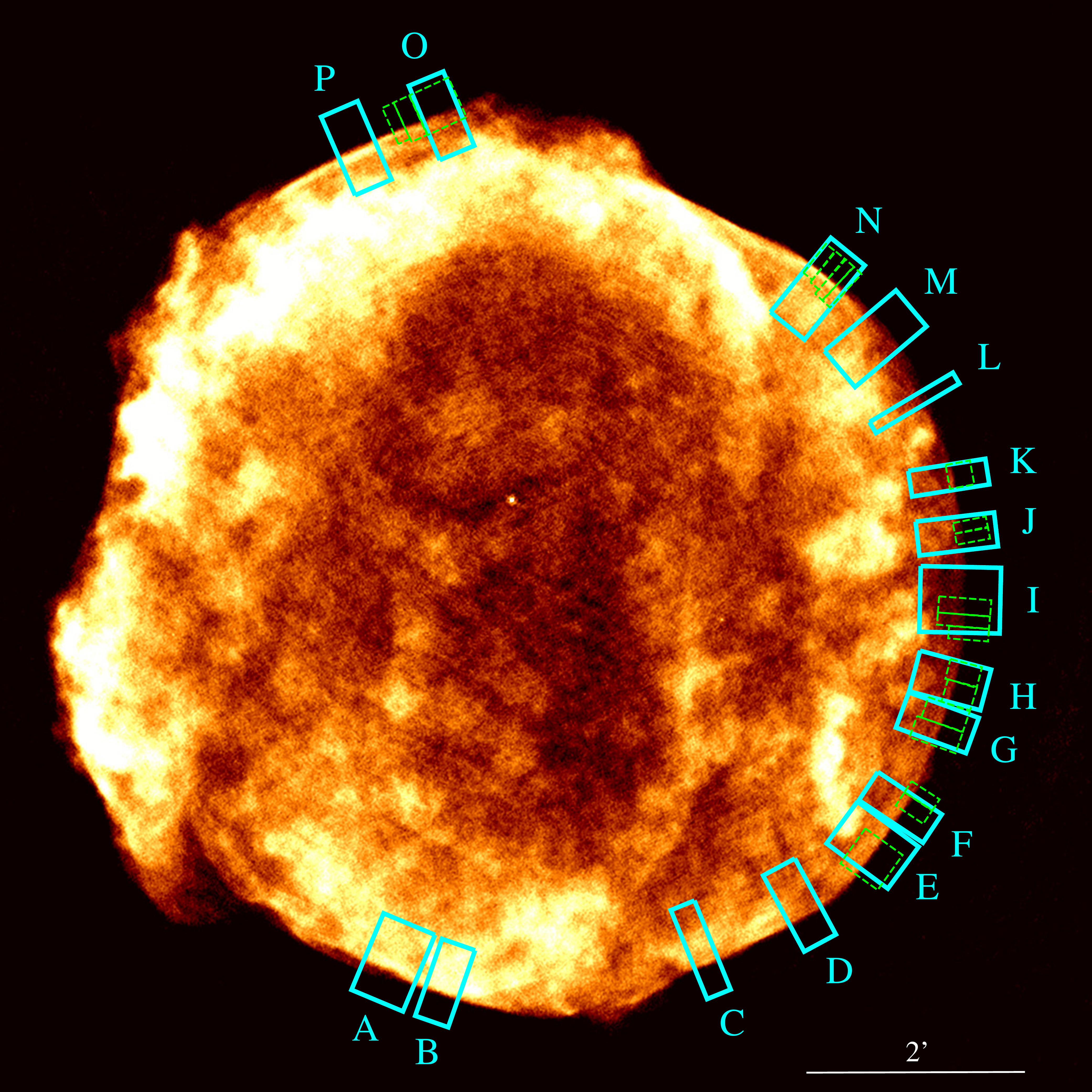}
    %\plotone{fig_radio-snr-inv.png}  % Save ink when printing
    \caption{Radio image of Tycho's SNR at $1.375 \unit{GHz}$ with linear
    scaling.  Extraction regions (green) for joint radio and X-ray profile
    analysis overlay region selections for X-ray rim width analysis
    (Figure~\ref{fig:snr}). \label{fig:radio-snr}}
\end{figure}

% Procedure: image/profiles
We extract radio profiles from a $1.375 \unit{GHz}$ image of Tycho taken
with the Very Large Array (VLA) in A configuration in
March 1994 (PI: D. Moffett); see \citet{reynoso1997} for a detailed
presentation.
The half-power beam width of $\abt 1.5\arcsec$ just resolves thin radio rims
and structure near the forward shock; the image is sampled at $0.5\arcsec$.
We also extracted $4$--$7 \unit{keV}$ X-ray profiles in all regions from the
previous archival \Chandra observation to jointly model radio and X-ray
profiles, permitting somewhat firmer discrimination of plausible model
parameters.  Figure~\ref{fig:radio-snr} shows the extraction regions overlaying
the radio image and the previous X-ray profile regions of Figure~\ref{fig:snr}.

% Procedure: model computation
We compute model radio and X-ray measured profiles from the transport model of
equation~\eqref{eq:model} for varying $B_0$ and $a_b$, similar to those
shown in Figure~\ref{fig:pargrid}.  The parameters
$B_{\mt{min}} = 5 \muG$, $\eta_2 = 1$, and $\mu = 1$ are held fixed.
Diffusion, in particular, is negligible for modeled radio emission as particle
energies are 3 orders of magnitude lower than in X-ray.  Neglecting diffusion
also circumvents the unphysical assumption that $D(x) B^2(x)$ is constant,
which was invoked to obtain Green's function solutions to
equation~\eqref{eq:model}.

% Procedure: chi-by-eye "fitting"
We align each set of model profiles at some $B_0$ and $a_b$
to measurements by eye, varying relative amplitudes and
translations in radio and X-ray independently to best match the measured
profiles.
Although ``fitting'' by eye does not quantitatively bound model parameters,
we can find plausible values of $B_0$ and $a_b$ and estimate the importance of
magnetic damping throughout the remnant.
More involved nonlinear fitting may be unreliable as we cannot constrain
spatially heterogeneous radio emission within the remnant.
Moreover, our transport model neglects self-similar downstream evolution of
shocked plasma (decaying velocity, density, and magnetic field) and is
inaccurate further downstream than $\abt 10\%$ of the shock radius.

% Procedure: cont
The joint radio and X-ray profile modeling contrasts strongly with our previous
width-energy fitting from X-ray measurements alone, which neglected profile
shape in favor of more robust FWHM measurements.  Manually fitting profiles
allows us to consider radio and X-ray filaments that do not have well-defined
FWHMs, especially as radio rims do not fall below 50\% of the peak emission.
We can also use profiles from regions not previously considered due to the lack
of an X-ray FWHM, especially in softer ($0.7$--$4 \unit{keV}$) X-rays.

\begin{table}
    \scriptsize
    \centering
    \caption{Fit parameters from profile shape comparison
        \label{tab:fits-eyeball}}
    \input{tab-fits-eyeball.tex}
    \tablecomments{Damping lengths of $\infty$ indicate that a
        loss-limited fit is favored ($a_b > 10$\% of shock radius $r_s$).}
\end{table}

\begin{figure}
    \centering
    \iftoggle{manuscript}{
        \epsscale{0.6}
        \plotone{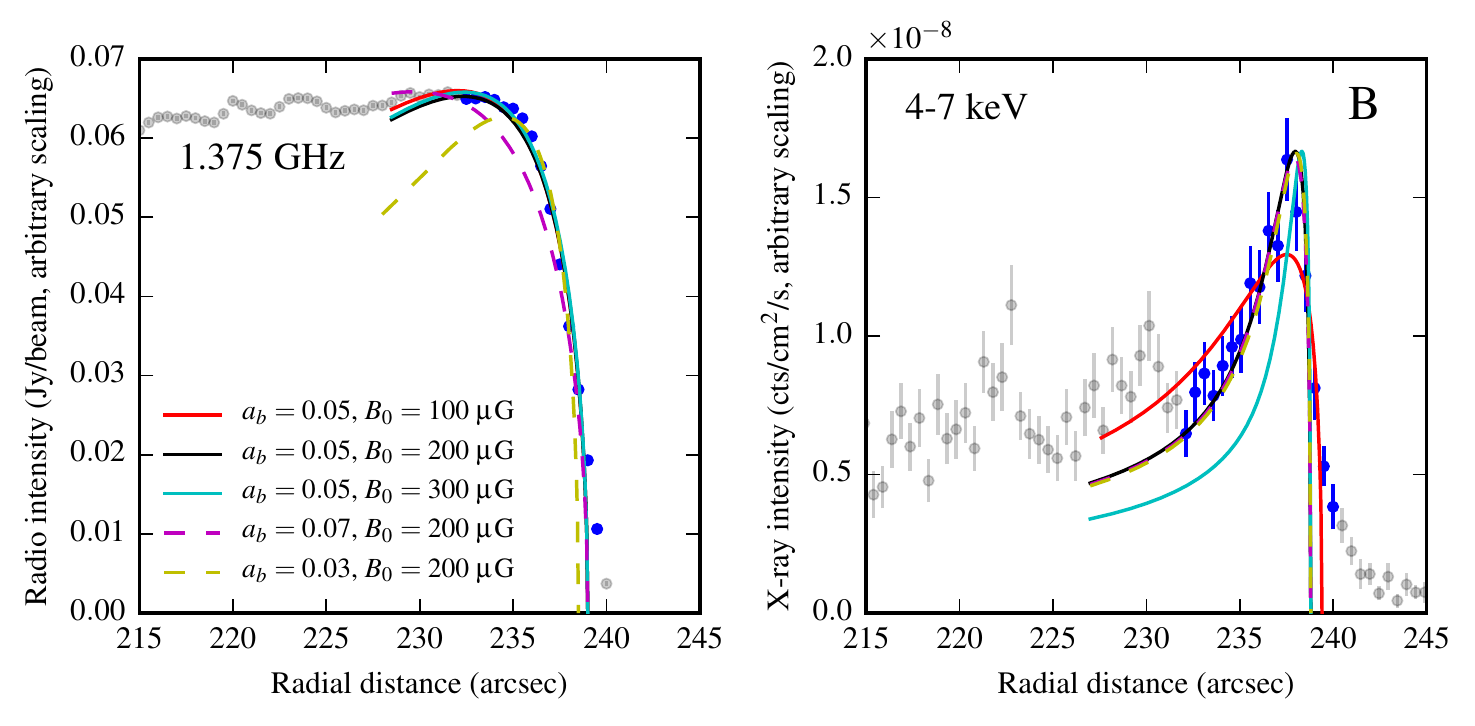} \\
        \plotone{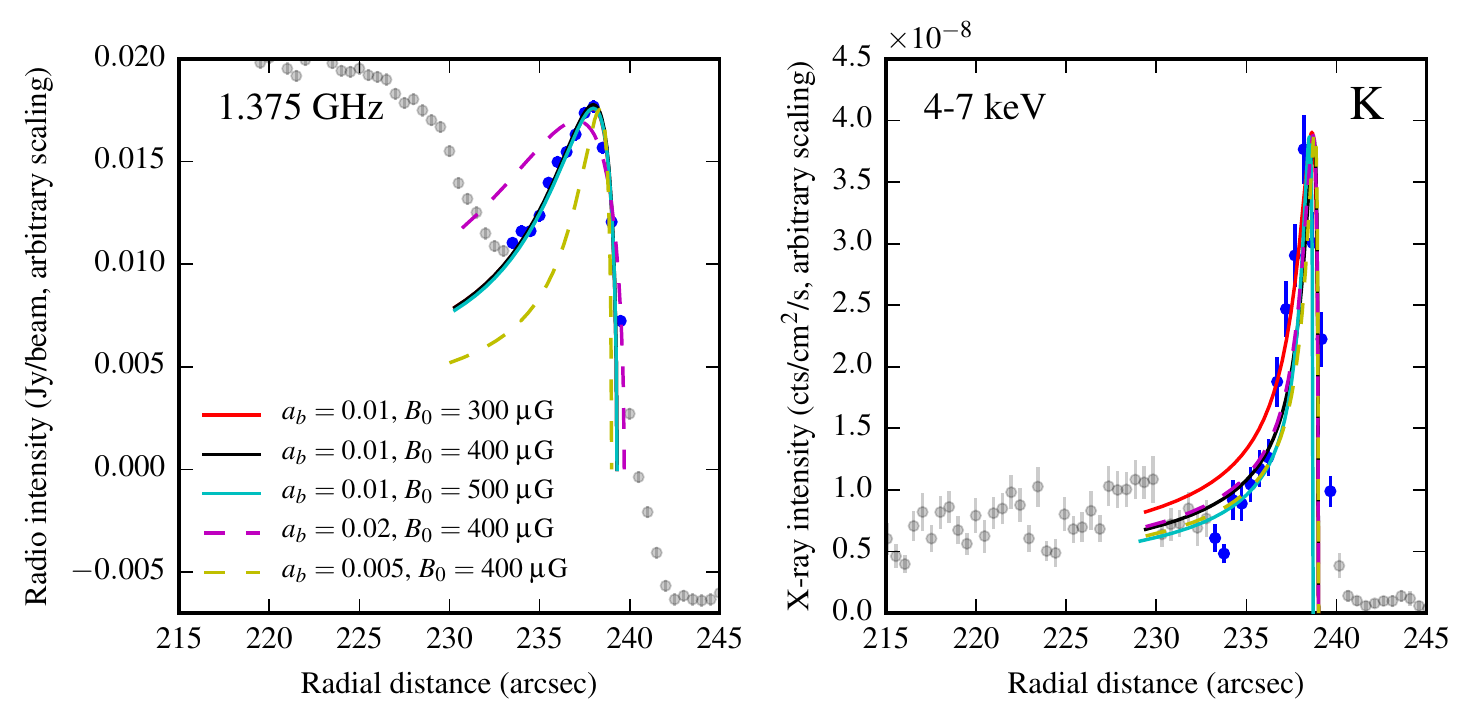} \\
        \plotone{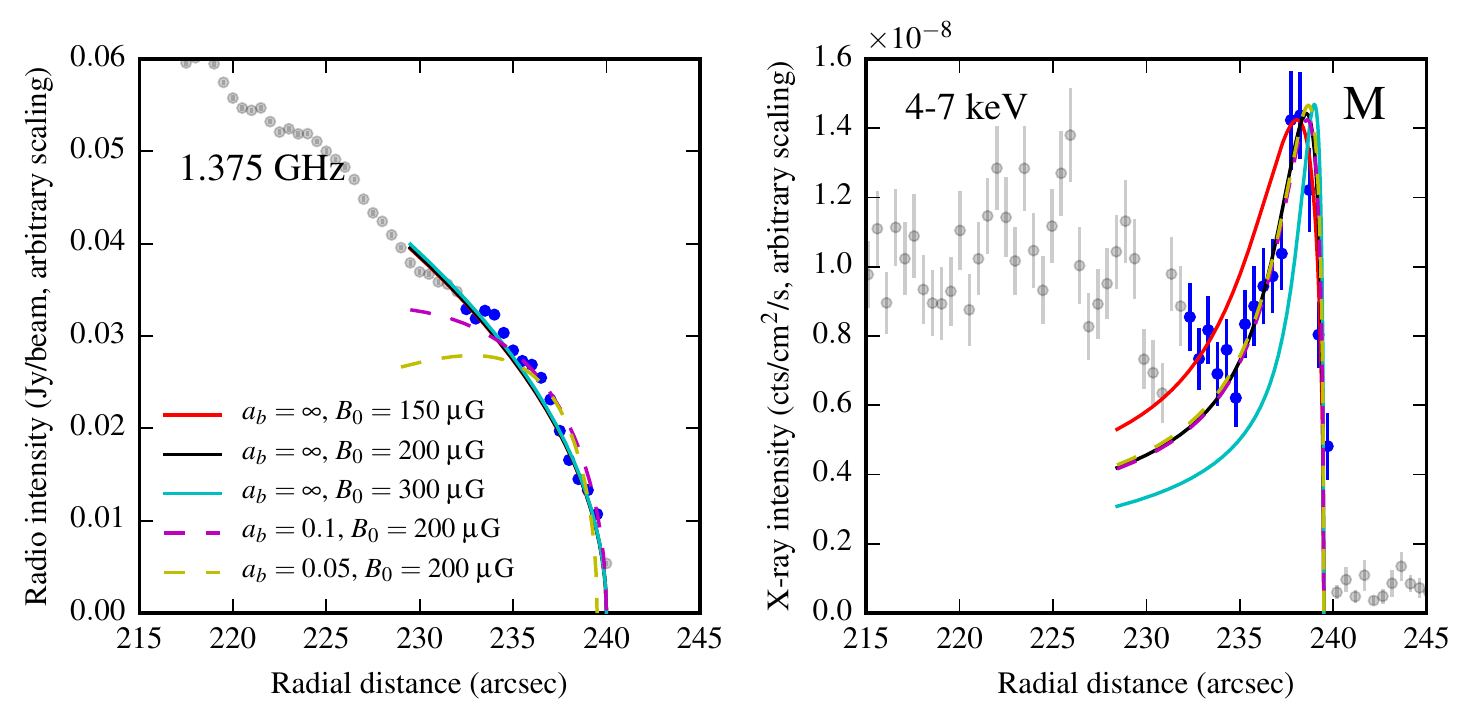}
        \epsscale{1}
    }{
        % Slightly larger figures than given by \plotone
        \includegraphics[width=0.46\textwidth]{fig_radio-fits-B.pdf} \\
        \includegraphics[width=0.46\textwidth]{fig_radio-fits-K.pdf} \\
        \includegraphics[width=0.46\textwidth]{fig_radio-fits-M.pdf}
    }
    \caption{Measured radio and X-ray profiles plotted with model profiles for
    varying $a_b$ and $B_0$ in each region, showing typical parameters (and
    ranges) required to reproduce radio and X-ray rims simultaneously in our
    model.  \emph{Solid black curves} in all regions plot our manually chosen best
    model profiles.  Profiles are chosen to show varying radio rim morphology,
    including plateaus (B), thin rims with troughs (K), and continuous rises
    (M); these profiles show some of the best agreement of our selected
    regions, but cf. Figure~\ref{fig:radio-prfs-meh}.  Profile radial
    coordinates are shifted arbitrarily to aid visual comparison.
    Negative radio intensity is unphysical and associated with deconvolution of
    raw VLA visibilities.
    \label{fig:radio-prfs}}
\end{figure}

% Results: morphology, individual regions, main figure
We identify three classes of radio profiles: thin rims with downstream troughs,
plateaus, and continuous rises.  Regions B, C, and D (southern limb) show plateaus
in radio emission.  Regions J, L, and M (around NW) show continuous rises in
emission.  All other regions have a radio rim within $15\arcsec$ of the forward
shock, where the forward shock in radio is assumed at zero intensity.
Figure~\ref{fig:radio-prfs} shows extracted radio and nonthermal X-ray
($4$--$7\unit{keV}$) profiles for three well-fit regions with our best manually
selected model profiles, illustrating each of the three radio profile types
observed.

\begin{figure}
    \centering
    \iftoggle{manuscript}{
        \plotone{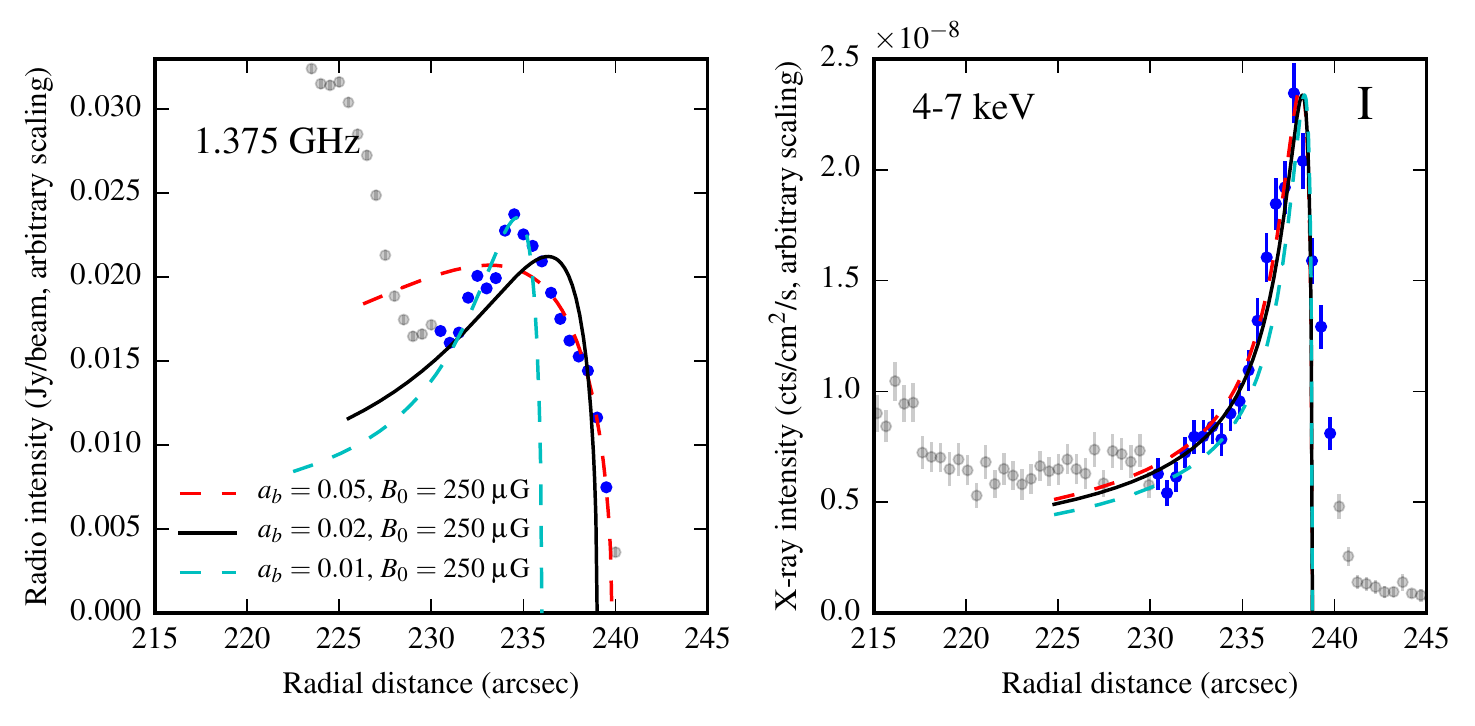} \\
        \plotone{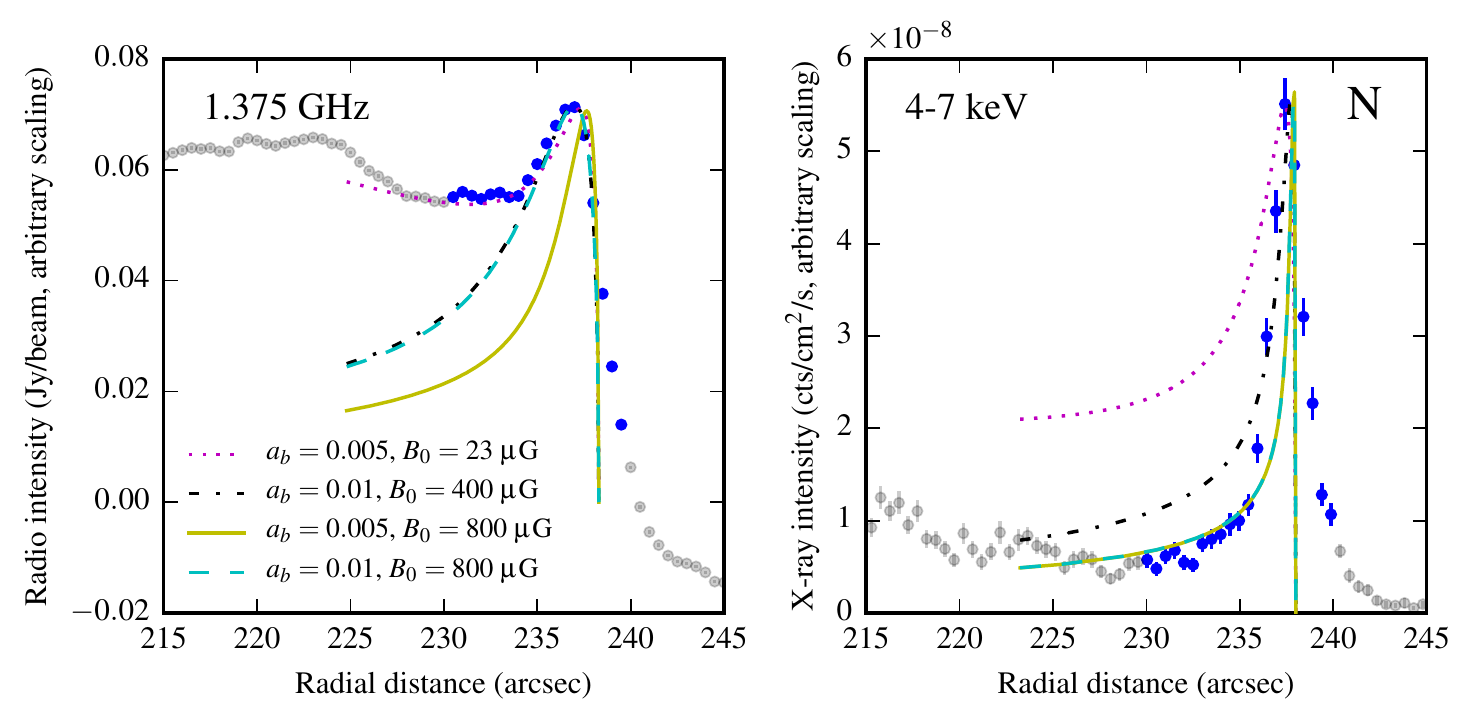}
    }{
        \includegraphics[width=0.47\textwidth]{fig_radio-fits-I.pdf} \\
        \includegraphics[width=0.47\textwidth]{fig_radio-fits-N.pdf}
    }
    \caption{Extracted profiles poorly reproduced by our model, compared
    to Figure~\ref{fig:radio-prfs}.  Region I shows irregularly shaped radio
    rim.  Region N contains two superposed filaments that cannot be modeled by
    a single rim in both X-ray and radio; the narrow X-ray rim requires
    atypically strong magnetic fields ($\sim 800 \muG$), and the emission
    plateau behind the radio rim requires small damping lengths ($a_b \sim
    0.005$).
    \label{fig:radio-prfs-meh}}
\end{figure}

% Results: linchpin for damping model here
Our model requires damping length $a_b \lesssim 0.1$ to produce a plateau or
thin rim in radio emission.
For regions with thin radio rims, the best manually selected profiles have
$B_0$ between $50$--$400 \muG$, neglecting only Region N which could not be
modeled simultaneously in both X-ray and radio (Figure~\ref{fig:radio-prfs-meh}).
The damping length $a_b$ ranges between $0.01$--$0.03$, or $2$--$7\arcsec$.
We list estimated best fit parameters in Table~\ref{tab:fits-eyeball}.

The radio plateaus in regions B, C, D are compatible with damping lengths
between $0.01$--$0.05$.  Continuous rises in radio emission are best modeled
with $a_b \gtrsim 0.1$ and $B_0 \abt 200$--$300 \muG$.  Although damping
lengths $a_b \gtrsim 0.1$ are not physically meaningful well beyond the shock,
these large values of $a_b$ yield practically constant magnetic field
near the shock.
Our results are also insensitive to the assumed value of $\Bmin$; only model
profiles with small magnetic fields ($B_0 \lesssim 100 \muG$) and/or small
damping length ($a_b < 0.01$) are affected if we take $\Bmin = 2 \muG$ rather
than $5 \muG$.

% Discussion: confounding effects (and figure), alternate explanations?
Tycho's shock structure is more complex than assumed by our transport model.
Emission towards the remnant interior clearly shows spatial structure
(Figure~\ref{fig:radio-snr}).  Figure \ref{fig:radio-prfs-meh} shows two
regions that were poorly described by our model.  A majority of our regions
have irregular rims; in at least 2--3 regions, this may be attributed to
projection of multiple filaments.
Others (e.g., Region K) show rims with slopes that cannot be matched by our
models, whether too steep or shallow.
Shape mismatch may be attributed in part to point-spread mismatch,
diffusion (e.g., $\eta_2 \neq 1$), shock precursors, or other effects.
Nevertheless, the conclusion that thin radio rims require magnetic damping is
supported by more sophisticated modeling.  As we have noted, hydrodynamic
models with diffusive shock acceleration \citep{cassam-chenai2007, slane2014}
also cannot produce radio profiles with narrow rims in a purely advected
magnetic field.
It is also not clear that radio structure is due to a global radial
variation in magnetic field strength.
E.g., \citet{slane2014} suggest that Rayleigh-Taylor fingers between the
forward shock and contact discontinuity could locally confine radio-emitting
electrons, creating observed radio rim structure.
But, if radio rims are due strictly to magnetic field effects, our conclusions
on magnetic field drop-off are independent of the causative physical mechanism
so long as they yield magnetic field fall-off over 1\% of remnant radius.

% Back to X-ray width-energy fits -- weak-field damping seems unlikely with
% these radio rims.
We may also use spatially overlapping radio and X-ray region selections
(Figure~\ref{fig:radio-snr}) to attempt to constrain X-ray width-energy fits.
All X-ray region selections (Figure~\ref{fig:snr}) are associated with radio
rims, except for Regions 11 and 12 underlying Region J in radio.
X-ray emission at regions B, C, D (radio plateaus) and L, M (radio rises)
showed rims, but their widths either could not be measured, or could only be
measured in the $4$--$7 \unit{keV}$ band.
If radio rims require damping at $\abt1\%$ remnant radius, width-energy
fits with ``weak-field'' damping lengthscales of $\abt0.5\%$ may be disfavored.
Conversely, the radio and X-ray filament of Regions 11,12 and J is the best
remaining loss-limited rim candidate, although width-energy fits are equivocal
towards damped and loss-limited models.

% ==========
% Conclusion
% ==========
\section{Conclusions}

We measured the widths of several thin synchrotron filaments around Tycho's
supernova remnant and found moderate narrowing of rim widths throughout the
remnant, corroborating rim narrowing observed by \citet{ressler2014} in the
remnant of SN 1006.  We confirmed that selected filaments are dominated by
nonthermal emission and have clearly measurable full widths at half maximum in
4--5 energy bands.
Both X-ray width-energy fits and joint radio/X-ray profile modeling
require magnetic fields $\gtrsim 20 \muG$ even with magnetic damping.

A steady-state particle transport model with constant magnetic field gives
diffusion coefficients and magnetic field strengths broadly consistent with
prior estimates from rim widths \citep[e.g.,][]{parizot2006, rettig2012} and
radio and gamma ray measurements \citep{acciari2011, morlino2012}.  The same
model with a damped magnetic field is equally capable of describing our
measured data.  At weak energy dependence the two models are indistinguishable
and magnetic damping fits favor moderately amplified magnetic
fields beyond simple compression, but lower than for loss-limited models.
At moderate energy dependence ($\mE \sim -0.3$), the damping model permits weak
magnetic fields and short damping lengths ($<1\%$ remnant radius) to reproduce
energy dependence, but we still cannot favor either damped or loss-limited
rims due to fit uncertainty.
The distinction between loss-limited and damped models is
somewhat artificial; only for large differences between the various transport
lengthscales, a frequency-dependent occurrence, is one or the other mechanism
clearly dominant.

Thin radio synchrotron rims, however, are not reproduced in loss-limited models
\citep{cassam-chenai2007}.
Assuming shocked electrons account for most radio emission immediately
downstream of Tycho's forward shock, we jointly model radio and X-ray profiles
and find that damping lengths of $1$--$5$\%
of the shock radius are required throughout most of the remnant; only a few
(3/16) selected regions are plausibly consistent with a constant advected
magnetic field.
Typical magnetic field strengths range between $50$--$400 \muG$.
Although we cannot bound damping lengths and fields from our qualitative
profile comparisons, our results are physically reasonable and are likely good
to order-of-magnitude.
If damping lengths inferred from radio rims are correct, ``weak-field'' damping
is disfavored in explaining X-ray rim width-energy dependence, and damped rims
require magnetic field amplification to $\abt 100 \muG$ or more
in Tycho.

% ================
% Acknowledgements
% ================
\acknowledgments

We thank the anonymous referee for comments that helped clarify and
improve the manuscript.
David Moffett kindly provided the VLA image used in this study.
The scientific results reported in this article are based on data obtained from
the \Chandra Data Archive.
This research made extensive use of NASA's Astrophysics Data System.
This research also made use of APLpy, an open-source plotting package for
Python hosted at \href{http://aplpy.github.com}{http://aplpy.github.com}.

{\it Facilities:} \facility{CXO (ACIS-I)}, \facility{VLA}

% ==========
% References
% ==========
\bibliographystyle{apj}  % AASTeX journal macros are supplied in ADS entries
\bibliography{refs-snr}

% ========
% Appendix
% ========
%\clearpage
%\newpage
%\appendix
%
%\setcounter{table}{0}
%\renewcommand{\thetable}{A\arabic{table}}
%\setcounter{figure}{0}
%\renewcommand{\thefigure}{A\arabic{figure}}

\end{CJK}
\end{document}

%% file: tab-spec.tex
\begin{tabular}{@{}lcclccrccr@{}}
\toprule
{} & \multicolumn{3}{c}{Downstream, power-law}
   & \multicolumn{3}{c}{Rim, power-law}
   & \multicolumn{3}{c}{Rim, \texttt{srcut}} \\
\cmidrule(lr){2-4} \cmidrule(lr){5-7} \cmidrule(l){8-10}
Region & $N_{\mt{H}}$ & $\Gamma$ & $\chi^2_{\mathrm{red}}$ (dof)
       & $N_{\mt{H}}$ & $\Gamma$ & $\chi^2_{\mathrm{red}}$ (dof)
       & $N_{\mt{H}}$ & $\nu_{\mt{cut}}$ & $\chi^2_{\mathrm{red}}$ (dof) \\
{} & ($10^{22} \mt{cm}^{-2}$) & (-) & {}
   & ($10^{22} \mt{cm}^{-2}$) & (-) & {}
   & ($10^{22} \mt{cm}^{-2}$) & (keV/$h$) & {} \\
\midrule
 1 & 0.53 & 2.72 & 2.25 (186) & 0.72 & 2.84 & 1.20 (284)
   & 0.62 & 0.30 & 1.17 (284) \\
 2 & 0.68 & 2.99 & 5.07 (178) & 0.69 & 2.83 & 1.10 (202)
   & 0.60 & 0.30 & 1.08 (202) \\
 3 & 0.67 & 2.96 & 1.97 (186) & 0.77 & 2.80 & 1.15 (167)
   & 0.68 & 0.33 & 1.14 (167) \\
\cmidrule{1-10}
 4 & 0.59 & 2.97 & 1.43 (163) & 0.70 & 2.84 & 1.21 (278)
   & 0.61 & 0.29 & 1.15 (278) \\
 5 & 0.62 & 2.93 & 4.76 (265) & 0.73 & 2.88 & 1.18 (255)
   & 0.64 & 0.27 & 1.11 (255) \\
 6 & 0.68 & 3.00 & 2.12 (200) & 0.74 & 2.85 & 0.96 (231)
   & 0.64 & 0.29 & 0.93 (231) \\
 7 & 0.65 & 3.02 & 1.00 (142) & 0.82 & 2.97 & 1.14 (224)
   & 0.71 & 0.23 & 1.14 (224) \\
 8 & 0.74 & 2.93 & 1.38 (170) & 0.75 & 2.71 & 0.98 (198)
   & 0.66 & 0.41 & 0.96 (198) \\
 9 & 0.78 & 3.03 & 1.12 (157) & 0.82 & 2.83 & 0.90 (175)
   & 0.73 & 0.30 & 0.88 (175) \\
10 & 0.62 & 2.86 & 1.40 (220) & 0.77 & 2.76 & 0.98 (164)
   & 0.68 & 0.36 & 0.96 (164) \\
\cmidrule{1-10}
11 & 0.67 & 2.94 & 2.56 (137) & 0.69 & 2.60 & 1.10 (153)
   & 0.61 & 0.55 & 1.07 (153) \\
12 & 0.61 & 2.79 & 2.65 (137) & 0.64 & 2.44 & 0.90 (172)
   & 0.57 & 0.88 & 0.90 (172) \\
13 & 0.61 & 2.98 & 3.12 (198) & 0.67 & 2.73 & 1.12 (235)
   & 0.59 & 0.38 & 1.09 (235) \\
\cmidrule{1-10}
14 & 0.46 & 2.93 & 1.37 (148) & 0.63 & 2.93 & 0.96 (167)
   & 0.54 & 0.23 & 0.95 (167) \\
15 & 0.43 & 2.92 & 1.33 (150) & 0.65 & 2.84 & 1.05 (183)
   & 0.57 & 0.29 & 1.03 (183) \\
16 & 0.48 & 2.94 & 2.04 (189) & 0.67 & 2.80 & 1.13 (182)
   & 0.58 & 0.32 & 1.12 (182) \\
17 & 0.48 & 2.86 & 1.70 (188) & 0.68 & 2.83 & 0.97 (187)
   & 0.59 & 0.30 & 0.94 (187) \\
\cmidrule{1-10}
18 & 0.44 & 2.87 & 1.91 (200) & 0.64 & 3.02 & 1.20 (220)
   & 0.55 & 0.19 & 1.13 (220) \\
19 & 0.40 & 2.84 & 1.31 (133) & 0.66 & 2.78 & 1.01 (157)
   & 0.57 & 0.34 & 0.99 (157) \\
20 & 0.40 & 2.75 & 3.01 (140) & 0.63 & 2.81 & 1.11 (192)
   & 0.55 & 0.31 & 1.07 (192) \\
\bottomrule
\end{tabular}

%% file: tab-fwhms.tex
\begin{tabular}{@{}l ccccc r@{ $\pm$ }l r@{ $\pm$ }l r@{ $\pm$ }l r@{ $\pm$ }l @{}}

\toprule
{} & \multicolumn{5}{c}{FWHM (arcsec)} & \multicolumn{8}{c}{$\mE$ (-)} \\
\cmidrule(lr){2-6} \cmidrule(l){7-14}
Region & Band 1 & Band 2 & Band 3 & Band 4 & Band 5
       & \multicolumn{2}{c}{Bands 1--2} & \multicolumn{2}{c}{Bands 2--3}
       & \multicolumn{2}{c}{Bands 3--4} & \multicolumn{2}{r}{Bands 4--5} \\ [0.2em]
{} & (0.7--1 keV) & (1--1.7 keV) & (2--3 keV) & (3--4.5 keV) & (4.5--7 keV)
   & \multicolumn{2}{c}{(1 keV)} & \multicolumn{2}{c}{(2 keV)}
   & \multicolumn{2}{c}{(3 keV)} & \multicolumn{2}{r}{(4.5 keV)} \\
\midrule
1 & {} & ${8.80}^{+0.18}_{-0.15}$ & ${6.34}^{+0.26}_{-0.21}$ & ${7.40}^{+0.30}_{-0.23}$ & ${5.57}^{+0.47}_{-0.42}$
  & \multicolumn{2}{c}{} & $-0.47$ & $0.06$ & $0.38$ & $0.13$ & $-0.70$ & $0.22$ \\ [0.5em]
2 & {} & ${4.22}^{+0.12}_{-0.09}$ & ${2.36}^{+0.12}_{-0.09}$ & ${3.00}^{+0.16}_{-0.12}$ & ${4.11}^{+0.34}_{-0.30}$
  & \multicolumn{2}{c}{} & $-0.84$ & $0.08$ & $0.59$ & $0.16$ & $0.77$ & $0.23$ \\ [0.5em]
3 & {} & ${2.47}^{+0.08}_{-0.07}$ & ${1.78}^{+0.09}_{-0.07}$ & ${2.10}^{+0.11}_{-0.11}$ & ${1.32}^{+0.10}_{-0.09}$
  & \multicolumn{2}{c}{} & $-0.47$ & $0.08$ & $0.41$ & $0.17$ & $-1.15$ & $0.22$ \\

\cmidrule{1-14}
4 & ${5.85}^{+0.37}_{-0.33}$ & ${4.35}^{+0.09}_{-0.08}$ & ${3.26}^{+0.11}_{-0.09}$ & ${3.69}^{+0.12}_{-0.11}$ & ${3.20}^{+0.21}_{-0.18}$
  & $-0.83$ & $0.18$ & $-0.41$ & $0.05$ & $0.31$ & $0.11$ & $-0.35$ & $0.17$ \\ [0.5em]
5 & {} & ${4.52}^{+0.11}_{-0.12}$ & ${3.06}^{+0.11}_{-0.11}$ & ${3.25}^{+0.15}_{-0.13}$ & ${3.04}^{+0.21}_{-0.18}$
  & \multicolumn{2}{c}{} & $-0.56$ & $0.06$ & $0.15$ & $0.14$ & $-0.17$ & $0.19$ \\ [0.5em]
6 & ${2.48}^{+0.18}_{-0.18}$ & ${2.32}^{+0.05}_{-0.06}$ & ${2.98}^{+0.11}_{-0.09}$ & ${2.05}^{+0.08}_{-0.09}$ & ${2.21}^{+0.15}_{-0.14}$
  & $-0.19$ & $0.21$ & $0.36$ & $0.06$ & $-0.92$ & $0.13$ & $0.18$ & $0.19$ \\ [0.5em]
7 & ${2.69}^{+0.20}_{-0.17}$ & ${2.33}^{+0.05}_{-0.05}$ & ${2.31}^{+0.08}_{-0.08}$ & ${1.81}^{+0.09}_{-0.07}$ & ${1.83}^{+0.11}_{-0.08}$
  & $-0.39$ & $0.20$ & $-0.01$ & $0.06$ & $-0.60$ & $0.14$ & $0.02$ & $0.17$ \\ [0.5em]
8 & ${2.33}^{+0.21}_{-0.20}$ & ${2.72}^{+0.08}_{-0.08}$ & ${2.38}^{+0.10}_{-0.09}$ & ${2.10}^{+0.10}_{-0.09}$ & ${2.37}^{+0.20}_{-0.17}$
  & $0.43$ & $0.26$ & $-0.19$ & $0.07$ & $-0.30$ & $0.15$ & $0.29$ & $0.22$ \\ [0.5em]
9 & ${2.16}^{+0.24}_{-0.23}$ & ${2.35}^{+0.07}_{-0.06}$ & ${2.47}^{+0.11}_{-0.11}$ & ${1.91}^{+0.09}_{-0.09}$ & ${2.20}^{+0.17}_{-0.16}$
  & $0.24$ & $0.31$ & $0.07$ & $0.07$ & $-0.63$ & $0.16$ & $0.34$ & $0.22$ \\ [0.5em]
10 & ${2.38}^{+0.24}_{-0.23}$ & ${1.99}^{+0.07}_{-0.06}$ & ${1.76}^{+0.09}_{-0.08}$ & ${1.59}^{+0.09}_{-0.08}$ & ${1.58}^{+0.13}_{-0.12}$
  & $-0.50$ & $0.29$ & $-0.18$ & $0.08$ & $-0.24$ & $0.18$ & $-0.02$ & $0.23$ \\

\cmidrule{1-14}
11 & {} & ${3.23}^{+0.15}_{-0.13}$ & ${2.52}^{+0.16}_{-0.13}$ & ${1.90}^{+0.14}_{-0.13}$ & ${3.09}^{+0.45}_{-0.38}$
  & \multicolumn{2}{c}{} & $-0.36$ & $0.10$ & $-0.70$ & $0.22$ & $1.21$ & $0.37$ \\ [0.5em]
12 & {} & ${3.86}^{+0.17}_{-0.16}$ & ${2.61}^{+0.15}_{-0.13}$ & ${3.02}^{+0.22}_{-0.21}$ & ${2.23}^{+0.21}_{-0.17}$
  & \multicolumn{2}{c}{} & $-0.56$ & $0.10$ & $0.36$ & $0.22$ & $-0.74$ & $0.27$ \\ [0.5em]
13 & ${2.85}^{+0.22}_{-0.17}$ & ${2.43}^{+0.05}_{-0.05}$ & ${2.36}^{+0.08}_{-0.05}$ & ${1.95}^{+0.09}_{-0.10}$ & ${1.84}^{+0.11}_{-0.14}$
  & $-0.45$ & $0.20$ & $-0.04$ & $0.05$ & $-0.47$ & $0.13$ & $-0.15$ & $0.20$ \\

\cmidrule{1-14}
14 & ${2.86}^{+0.17}_{-0.16}$ & ${2.42}^{+0.06}_{-0.04}$ & ${2.23}^{+0.08}_{-0.07}$ & ${2.38}^{+0.10}_{-0.08}$ & ${2.19}^{+0.12}_{-0.10}$
  & $-0.47$ & $0.17$ & $-0.12$ & $0.06$ & $0.17$ & $0.12$ & $-0.20$ & $0.15$ \\ [0.5em]
15 & ${2.71}^{+0.17}_{-0.16}$ & ${1.99}^{+0.05}_{-0.04}$ & ${1.80}^{+0.06}_{-0.05}$ & ${1.87}^{+0.07}_{-0.05}$ & ${1.52}^{+0.09}_{-0.08}$
  & $-0.85$ & $0.18$ & $-0.15$ & $0.05$ & $0.09$ & $0.11$ & $-0.51$ & $0.16$ \\ [0.5em]
16 & ${1.87}^{+0.14}_{-0.13}$ & ${1.73}^{+0.04}_{-0.03}$ & ${1.52}^{+0.06}_{-0.05}$ & ${1.25}^{+0.06}_{-0.04}$ & ${1.23}^{+0.08}_{-0.06}$
  & $-0.22$ & $0.21$ & $-0.18$ & $0.06$ & $-0.49$ & $0.13$ & $-0.04$ & $0.17$ \\ [0.5em]
17 & ${1.65}^{+0.13}_{-0.12}$ & ${1.92}^{+0.05}_{-0.05}$ & ${1.54}^{+0.06}_{-0.07}$ & ${1.45}^{+0.07}_{-0.06}$ & ${2.05}^{+0.16}_{-0.14}$
  & $0.43$ & $0.22$ & $-0.31$ & $0.07$ & $-0.16$ & $0.15$ & $0.86$ & $0.21$ \\

\cmidrule{1-14}
18 & {} & ${4.45}^{+0.13}_{-0.12}$ & ${3.18}^{+0.17}_{-0.16}$ & ${2.96}^{+0.20}_{-0.19}$ & ${1.65}^{+0.21}_{-0.16}$
  & \multicolumn{2}{c}{} & $-0.49$ & $0.09$ & $-0.17$ & $0.21$ & $-1.45$ & $0.32$ \\ [0.5em]
19 & {} & ${2.30}^{+0.08}_{-0.06}$ & ${2.28}^{+0.11}_{-0.08}$ & ${2.16}^{+0.12}_{-0.11}$ & ${1.60}^{+0.17}_{-0.14}$
  & \multicolumn{2}{c}{} & $-0.02$ & $0.08$ & $-0.13$ & $0.17$ & $-0.74$ & $0.27$ \\ [0.5em]
20 & ${4.81}^{+0.31}_{-0.31}$ & ${1.84}^{+0.06}_{-0.03}$ & ${1.87}^{+0.08}_{-0.06}$ & ${1.56}^{+0.07}_{-0.06}$ & ${2.14}^{+0.23}_{-0.23}$
  & $-2.68$ & $0.19$ & $0.02$ & $0.07$ & $-0.44$ & $0.14$ & $0.77$ & $0.28$ \\

\midrule
Mean & $2.89 \pm 0.35$ & $3.11 \pm 0.37$ & $2.53 \pm 0.23$ & $2.47 \pm 0.30$ & $2.35 \pm 0.23$
  & $-0.46$ & $0.24$ & $-0.25$ & $0.06$ & $-0.14$ & $0.10$ & $-0.09$ & $0.15$ \\

\bottomrule
\end{tabular}
\tablecomments{Mean values computed for all regions; mean $\mE$ values are
averages for region $\mE$ values (i.e., not computed from mean FWHMs).  Errors
on mean values are standard errors of the mean.  Horizontal rules group
individual regions into filaments.}

%% file: tab-fits-all-eta2one.tex
\begin{tabular}{@{} l rr rrr @{}}
\toprule
{} & \multicolumn{2}{c}{Loss-limited}
   & \multicolumn{3}{c}{Damped} \\
\cmidrule(lr){2-3} \cmidrule(l){4-6}
Region & $B_0$ ($\mu$G) & $\chi^2$
       & $B_0$ ($\mu$G) & $\chi^2$ & $a_b$ \\
\midrule
 1 & 182 &  35.1 &  28 & 24.3 & 0.008 \\
 2 & 312 &  80.9 &  25 & 74.4 & 0.003 \\
 3 & 426 &  21.5 &  27 & 23.9 & 0.002 \\
\cmidrule{1-6}
 4 & 284 &  50.9 &  28 & 29.9 & 0.004 \\
 5 & 288 &  19.3 &  25 & 14.5 & 0.003 \\
 6 & 410 & 173.7 &  68 & 65.2 & 0.004 \\
 7 & 418 &  50.9 & 291 & 11.6 & 0.006 \\
 8 & 388 &  47.7 & 138 & 10.0 & 0.005 \\
 9 & 414 &  65.1 &  78 & 14.2 & 0.004 \\
10 & 466 &  17.1 &  29 &  1.5 & 0.002 \\
\cmidrule{1-6}
11 & 355 &  12.0 & 317 & 15.6 & 0.010 \\
12 & 317 &   9.6 &  26 & 10.4 & 0.003 \\
13 & 400 &  52.7 & 258 & 10.4 & 0.006 \\
\cmidrule{1-6}
14 & 383 & 102.9 &  60 & 12.7 & 0.004 \\
15 & 431 &  70.3 &  48 & 20.0 & 0.003 \\
16 & 493 &  15.0 & 434 &  4.0 & 0.006 \\
17 & 467 &  61.6 & 232 & 29.0 & 0.004 \\
\cmidrule{1-6}
18 & 283 &  28.1 &  24 & 27.5 & 0.003 \\
19 & 401 &  36.9 &  78 &  8.3 & 0.004 \\
20 & 463 & 121.8 &  58 & 93.8 & 0.003 \\
\bottomrule
\end{tabular}

%% file: tab-fits-all-srcutlog.tex
\begin{tabular}{@{} l rrr rrr @{}}
\toprule
{} & {}
   & \multicolumn{2}{c}{Loss-limited}
   & \multicolumn{3}{c}{Damped} \\
\cmidrule(lr){3-4} \cmidrule(l){5-7}
Region & $\eta_2$
       & $B_0$ ($\mu$G) & $\chi^2$
       & $B_0$ ($\mu$G) & $\chi^2$ & $a_b$ \\
\midrule
 1 & 12.0 & 256 & 29.1 &  19 & 25.5 & 0.008 \\
 2 & 11.8 & 443 & 99.4 &  18 & 72.0 & 0.003 \\
 3 & 10.6 & 599 & 31.8 &  19 & 29.3 & 0.002 \\
\cmidrule{1-7}
 4 & 12.0 & 402 & 36.2 &  19 & 34.5 & 0.004 \\
 5 & 12.9 & 419 & 32.5 &  19 & 19.1 & 0.004 \\
 6 & 11.9 & 568 & 89.7 & 147 & 61.9 & 0.006 \\
 7 & 15.3 & 617 & 11.7 & 473 & 11.1 & 0.009 \\
 8 &  8.4 & 511 & 15.1 & 333 & 10.0 & 0.008 \\
 9 & 11.2 & 573 & 23.3 &  66 & 13.7 & 0.005 \\
10 &  9.4 & 631 &  1.4 & 478 &  2.0 & 0.007 \\
\cmidrule{1-7}
11 &  6.1 & 452 & 18.1 &  21 & 17.5 & 0.003 \\
12 &  3.7 & 375 & 12.6 &  21 &  9.7 & 0.003 \\
13 &  8.4 & 529 & 13.8 & 368 &  9.9 & 0.008 \\
\cmidrule{1-7}
14 & 11.9 & 547 & 28.6 & 120 & 12.4 & 0.006 \\
15 &  9.5 & 598 & 23.1 & 426 & 20.0 & 0.007 \\
16 &  8.4 & 670 &  6.3 &  22 &  5.8 & 0.002 \\
17 &  9.0 & 637 & 32.6 &  32 & 29.4 & 0.003 \\
\cmidrule{1-7}
18 & 14.6 & 432 & 72.4 &  17 & 16.6 & 0.003 \\
19 &  8.9 & 542 & 12.8 & 209 &  7.7 & 0.006 \\
20 &  9.6 & 631 & 96.0 &  29 & 94.2 & 0.003 \\
\bottomrule
\end{tabular}

%% file: tab-lengthscale.tex
\begin{tabular}{@{}l cc cc rcrr@{}}
\toprule
{} & \multicolumn{2}{c}{Measurements}
   & \multicolumn{2}{c}{Loss-lim. fit}
   & \multicolumn{4}{c}{Damped fit} \\
\cmidrule(lr){2-3} \cmidrule(lr){4-5} \cmidrule(l){6-9}
Region & $v_d$ & $w$ (2 keV)
       & $l_{\mt{ad}}$ & $l_{\mt{ad}}/l_{\mt{diff}}$
       & $l_{\mt{ad}}$ & $a_b$ & $l_{\mt{ad}}/a_b$ & $w(2\unit{keV})/a_b$\\
{} & ($10^8 \unit{cm/s}$) & (\%$r_s$)
   & (\%$r_s$) & (-)
   & (\%$r_s$) & (\%$r_s$) & (-) & (-) \\
\midrule
 1 & 1.30 & 2.64 & 0.61 & 1.40 & 10.26 & 0.80 & 12.83 & 3.30 \\
 2 & 1.29 & 0.99 & 0.27 & 1.39 & 11.54 & 0.30 & 38.48 & 3.28 \\
 3 & 1.29 & 0.74 & 0.17 & 1.38 & 10.80 & 0.20 & 54.00 & 3.72 \\
\cmidrule{1-9}
 4 & 1.28 & 1.36 & 0.31 & 1.38 & 10.08 & 0.40 & 25.20 & 3.40 \\
 5 & 1.28 & 1.27 & 0.30 & 1.37 & 12.07 & 0.30 & 40.24 & 4.25 \\
 6 & 1.28 & 1.24 & 0.18 & 1.37 &  2.61 & 0.40 &  6.53 & 3.10 \\
 7 & 1.27 & 0.96 & 0.17 & 1.36 &  0.29 & 0.60 &  0.49 & 1.61 \\
 8 & 1.27 & 0.99 & 0.19 & 1.36 &  0.90 & 0.50 &  1.80 & 1.98 \\
 9 & 1.26 & 1.03 & 0.17 & 1.36 &  2.09 & 0.40 &  5.23 & 2.57 \\
10 & 1.26 & 0.73 & 0.14 & 1.35 &  9.21 & 0.20 & 46.05 & 3.66 \\
\cmidrule{1-9}
11 & 1.25 & 1.05 & 0.21 & 1.34 &  0.25 & 1.00 &  0.25 & 1.05 \\
12 & 1.24 & 1.09 & 0.25 & 1.33 & 10.80 & 0.30 & 35.98 & 3.62 \\
13 & 1.23 & 0.98 & 0.18 & 1.32 &  0.34 & 0.60 &  0.57 & 1.64 \\
\cmidrule{1-9}
14 & 1.14 & 0.93 & 0.17 & 1.23 &  2.84 & 0.40 &  7.10 & 2.32 \\
15 & 1.14 & 0.75 & 0.15 & 1.22 &  3.95 & 0.30 & 13.16 & 2.50 \\
16 & 1.13 & 0.63 & 0.12 & 1.21 &  0.14 & 0.60 &  0.24 & 1.06 \\
17 & 1.12 & 0.64 & 0.13 & 1.20 &  0.36 & 0.40 &  0.91 & 1.61 \\
\cmidrule{1-9}
18 & 1.15 & 1.32 & 0.28 & 1.23 & 11.29 & 0.30 & 37.64 & 4.42 \\
19 & 1.18 & 0.95 & 0.17 & 1.27 &  1.99 & 0.40 &  4.97 & 2.37 \\
20 & 1.17 & 0.78 & 0.14 & 1.26 &  3.08 & 0.30 & 10.28 & 2.59 \\
\bottomrule
\end{tabular}

%% file: tab-fits-eyeball.tex
\begin{tabular}{@{}lrr lrr@{}}
\toprule
%\multicolumn{5}{c}{Eyeballed fits} \\
%\midrule
% Region & $B_0$ & $a_b$ & $l_{\mt{ad}}(2\unit{keV})$ & $l_{\mt{ad}}/a_b$ \\
% {} & ($\mu$G) & (\%$r_s$) & (\%$r_s$) & (-) \\
Region & $B_0$ & $a_b$ &
Region & $B_0$ & $a_b$ \\
    {} & ($\mu$G) & (-) &
    {} & ($\mu$G) & (-) \\
\cmidrule(r){1-3} \cmidrule(l){4-6}
A &  50 & 0.020 & I & 250 & 0.020 \\
B & 200 & 0.050 & J & 300 & $\infty$ \\
C &  15 & 0.010 & K & 400 & 0.010 \\
D & 100 & 0.050 & L & 250 & $\infty$ \\
E & 120 & 0.030 & M & 200 & $\infty$ \\
F & 300 & 0.025 & N & 800 & 0.010 \\
G & 250 & 0.020 & O & 200 & 0.005 \\
H & 300 & 0.020 & P & 150 & 0.012 \\
\bottomrule
\end{tabular}